\documentclass[preprint,12pt]{elsarticle}
\usepackage[a4paper,top=2cm,bottom=2cm,left=2.5cm,right=2.5cm,marginparwidth=1.75cm]{geometry}
\usepackage{amssymb}
\usepackage{caption}
\usepackage{subfigure}
\usepackage{float}
\usepackage{hyperref}
\usepackage{amssymb}
\hypersetup{hidelinks,
	colorlinks=true,
	allcolors=black,
	pdfstartview=Fit,
	breaklinks=true}
\usepackage{amsmath}
\usepackage{graphicx}
\usepackage{pdfpages}
\usepackage{upgreek}
\usepackage[toc,page]{appendix}
\usepackage{url}
\journal{International Journal of Hydrogen Energy}

\begin{document}

\begin{frontmatter}

\title{Numerical simulation of two-phase flow in gas diffusion layer and gas channel of proton exchange membrane fuel cells}
\author[inst1]{Danan Yang \corref{mycorrespondingauthor}}
\cortext[mycorrespondingauthor]{Corresponding author}
\ead{danan.yang@energy.lth.se}
\author[inst1]{Himani Garg}
\affiliation[inst1]{organization={Department of Energy Sciences, Faculty of Engineering, Lund University},%Department and Organization
            addressline={P.O. Box 118}, 
            %city={},
            postcode={Lund, SE-221 00}, 
            %state={},
            country={Sweden}}
%\maketitle

\author[inst1]{Martin Andersson}

\begin{abstract}
Liquid water within the cathode Gas Diffusion Layer (GDL) and Gas Channel (GC) of Proton Exchange Membrane Fuel Cells (PEMFCs) is strongly coupled to gas transport properties, thereby affecting the electrochemical conversion rates. In this study, the GDL and GC regions are utilized as the simulation domain, which differs from previous studies that only focused on any one of them. A volume-of-fluid method is adopted to numerically investigate the two-phase flow (gas and liquid) behavior, e.g., water transport pattern evolution, water coverage ratio as well as local and total water saturation. To obtain GDL geometries, an in-house geometry-based method is developed for GDL reconstruction. Furthermore, to study the effect of GDL carbon fiber diameter, the same procedure is used to reconstruct three GDL structures by varying the carbon fiber diameter but keeping the porosity and geometric dimensions constant. The wall wettability is introduced with static contact angles at carbon fiber surfaces and channel walls. 
The results show that the GDL fiber microstructure has a significant impact on the two-phase flow patterns in the cathode field. 
Different stages of two-phase flow pattern evolution in both cathode domains are observed. The liquid water in the GDLs experiences water invasion, spreading, and rising, followed by the droplet breakthrough in the GDL/GC interface. In the GCs, the water droplets randomly experience accumulation, combination, attachment, and detachment. Due to the difference in wettability, the water coverage of the GDL/GC interface is smaller than that of the channel side and top walls.
It is also found that the water saturation inside the GDLs stabilizes after the water breakthrough, while local water saturation at the interface keeps irregular oscillations. 
Last but not the least, a water saturation balance requirement between the GDL and GC is observed. In terms of varying fiber diameter, a larger fiber diameter would result in less water saturation in the GDL but more water in the GC, in addition to faster water movement throughout the total domain.
\end{abstract}

\begin{graphicalabstract}
\begin{figure}[H]
\centering
\includegraphics[width=1\textwidth]{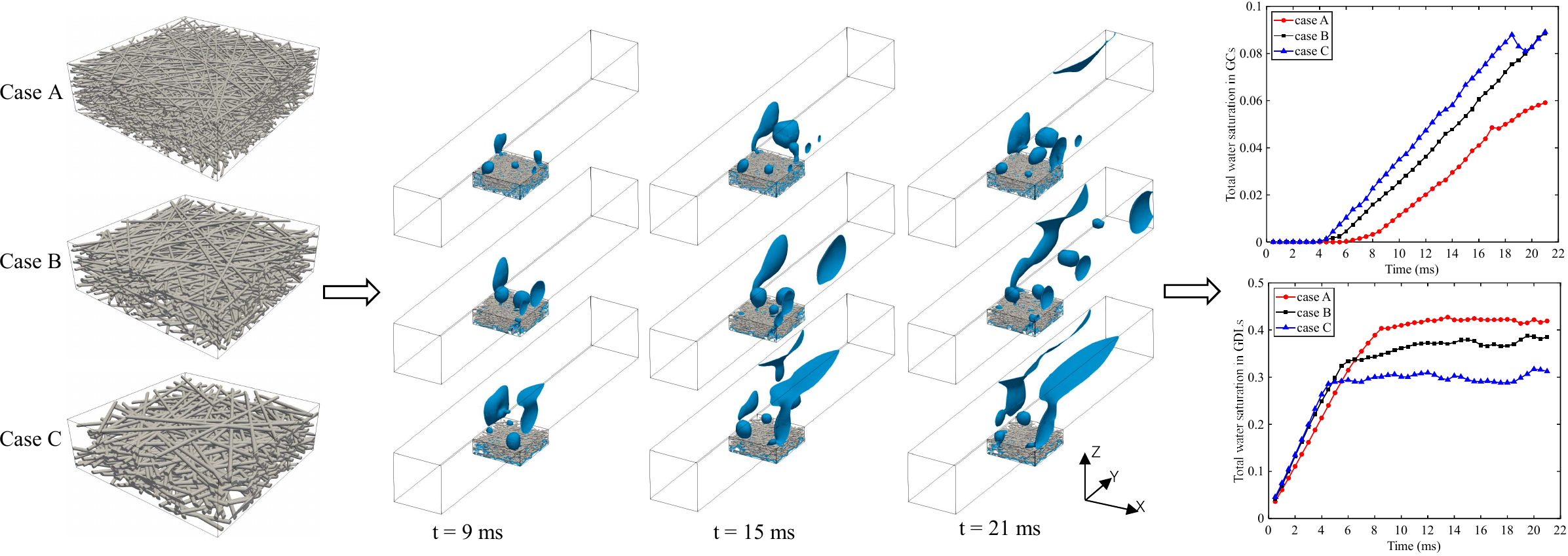}
\caption*{\label{fig:AbstractFig}Effect of GDL fiber diameter on water behavior in PEMFC gas channel and gas diffusion layer, varying the fiber diameters: 10 $\upmu$m (case A), 15 $\upmu$m (case B), and 20 $\upmu$m (case C).}
\end{figure}
\end{graphicalabstract}

\begin{highlights}
\item 
Using a volume-of-fluid method to study the two-phase flow in a cathode "T-shape" domain of PEMFCs.
\item Reconstruction of gas diffusion layers by an in-house geometry-based method.
\item Dynamic transport behavior of liquid water in the gas diffusion layer and gas channel.
\item Effect of gas diffusion layer fiber diameter on water transport.
\item Finding a water saturation balance requirement between the gas diffusion layer and the gas channel. 
\end{highlights}


\end{frontmatter}


\section{Introduction}

{Fuel cells have been considered to be an essential part of the future sustainable energy system. Proton Exchange Membrane Fuel Cells (PEMFCs) have attracted enough attention in the context of automotive, aerospace, and portable electronic devices because of their environmental friendliness and high power density \cite{peighambardoust2010review}. In a PEMFC, the Gas Diffusion Layers (GDLs) play a key role, especially the cathode one. It not only offers mechanical support for both side components but also provides fluid transport paths for gas diffusion and liquid water removal between the cathode Gas Channel (GC) and Proton Exchange Membrane (PEM). Water saturation in the cathode GDL and GC is really related to membrane humidity and gas transport performance. Without good water management, some bad operation situations might happen, for example, membrane dehydration results in increased resistive losses, while excessive amounts of water lead to flooding of the cathode domain, blocking the transport of reactants \cite{bao2006analysis}. In this regard, reactive gas starvation will weaken the reaction rate and even cause the degradation of materials. Moreover, a scientific understanding of water and thermal management \cite{pan2021review} and performance degradation \cite{r3} is not yet complete. Thus, an in-depth numerical investigation of the two-phase flow behavior inside the GDL and GC will be beneficial for understanding the water management of PEMFCs, enabling optimization of species transport properties, thereby improving the performance and decreasing the degradation.

{Before conducting a numerical study of the water transport mechanism within a GDL, the rather complex geometry can be reconstructed by either geometry-based stochastic generation algorithms \cite{zhu2021stochastically} or image-based structural scanning experimental data \cite{ r33}. A numerical stochastic reconstruction is based on random generation algorithms, simultaneously introducing structural parameters such as porosity and carbon fiber diameter. In contrast, image-based reconstruction combines X-ray Computed Tomography (CT) or Scanning Electron Microscope (SEM) slices from real geometries to rebuild the 3D geometries from a series of 2D slices. Zhu et al. \cite{zhu2021stochastically} reconstructed a carbon paper virtually to study the effect of binder and PolyteTraFluoroEthylene (PTFE) on effective gas diffusivity. Bosomoiu et al. \cite{r33} obtained fresh and aged gas diffusion layers by the X-ray CT reconstruction method, and studied the effective transport properties. Both stochastic and experimental reconstructions of GDLs have been done to analyze anisotropic transport properties in \cite{zhang2021microstructure}. Although a numerically stochastic reconstruction geometry does not completely agree with an actual structure, it saves time and cost within acceptable error tolerances, compared with experiment-based reconstruction approaches \cite{CHEN20218640}. Due to the attractive advantages of the geometry-based method, it becomes more and more popular in recent years and is selected for our present research.}

In the past years, to seek a balance between membrane drying and cathode flooding, plenty of research has been carried out for the two-phase flow in PEMFCs. Qin et al. \cite{CHEN20218640} reviewed the recent progress of GDL, focusing on visualization techniques of material structure and two-phase flow modeling as well as impacts of material properties. Andersson et al. \cite{andersson2016review} pointed out that full insight into the fundamental processes of liquid water evolution and transport in the GDL and GC is still lacking.

The research interests mainly focused on two-phase flow behavior in the GC or GDL separately. Due to the complex porous structure, the two-phase flow in the GDL seems to have attracted considerable attention. The topics were mainly related to the effect of GDL structure properties (types \cite{patel2019investigating}, compression and deformation \cite{BAO2021121608, chen2021numerical}, porosity \cite{guo2022pore}, wettability \cite{niu2018two}, capillary hysteresis \cite{sarkezi2022lattice}) and phase change phenomenon (ice-melting \cite{zhang2022study} and vapor condensation \cite{jiao2021vapor}). {Specifically, the effect of types was studied by Patel et al. \cite{patel2019investigating}, and water droplet characteristics in different GDLs were investigated by using synchrotron X-ray imaging technique to capture droplet dynamics. The GDL porous structure with a fiber microstructure that is highly sensitive to compression, gives a substantial change in materials and transport properties. As for the compression and deformation, based on stochastic orientation and finite element method, Bao et al. \cite{BAO2021121608} concluded that compression reduces the oxygen diffusivity and intrinsic permeability while it improves the thermal and electrical conductivity. Chen et al. \cite{chen2021numerical} developed a multi-physics and two-phase flow model to investigate the relationships between the non-uniform deformation and variation of physical properties of the GDL, as well as the cell performance. The effect of porosity was studied by Guo et al. \cite{guo2022pore} based on a novel pore-scale model. It was found that GDL with increased porosity from bottom to top was desirable. The GDL wettability impacts the two-phase flow behavior. PTFE is usually used to adjust the hydrophilic and hydrophobic properties of a GDL, controlling different water saturation. Niu et al. \cite{niu2018two} applied the Volume-Of-Fluid (VOF) method to investigate the effects of different PTFE loadings on capillary pressure and water saturation with PTFE distributions from two PTFE drying methods, i.e., vacuum- and air drying, respectively. For the phase change phenomenon, Zhang et al. \cite{zhang2022study} developed a gradient GDL ice melting model to investigate the ice-melting process; the effect of carbon fibers, i.e., growth slopes, numbers, and diameter size, on the GDL ice melting was studied. Jiao et al. \cite{jiao2021vapor} employed a novel VOF method to study the vapor/condensate water two-phase flow problem based on the stochastically reconstructed 3D GDL. Besides, a numerical stochastic reconstructed GDL and VOF method was used by Shi et al. \cite{SHI2021} to study the effect of the Micro-Porous Layer (MPL) cracks (shapes, numbers, and distance between cracks) on liquid water transport in the GDL. The results show that the GDL liquid water saturation positively correlates with crack number and the distance between cracks, while only a small correlation to the crack shape.} 

{In terms of two-phase flow in the GC of PEMFCs, it was numerically studied by placing single/multiple droplets on the GC/GDL interface or by injecting water into the channels from GDL. Most of the research aims to understand the removal time of water, drainage conditions, and the wettability of the GC/GDL surfaces. Specifically, two-phase flow distribution within a fuel cell multi-gas channel was studied using the VOF method, and the effects of operating conditions, wettability as well as microstructure were explored in \cite{zhang2022numerical}. The results showed that rectangular channels have better uniformity of flow distribution. In the previous study of Andersson et al. \cite{andersson2018modeling}, the effect of the static contact angle of the GC/GDL surface on the droplet detachment in GCs was explored with the VOF method, and experimental results from synchrotron-based X-ray radiography and tomography imaging were used for validation.
In addition, the dynamic contact angle boundary condition was also applied to the same geometry, resulting in different detachment times and droplet sizes, compared to static boundary conditions \cite{ANDERSSON201911088}. 

To our best knowledge so far, two-phase flow behavior in the connected GDL and GC of PEMFCs is rarely studied. There is still a lack of a complete investigation of the water removal process in the cathode field of the fuel cells. In addition, there are still few studies on the effects of wide variations in fiber diameter. Thus, the two-phase flow study in the GC only, by Andersson et al. \cite{ANDERSSON201911088,ANDERSSON20182961}, is now extended to investigate the two-phase flow phenomena in both CGL and GDL as well as their coupled interfaces. simultaneously. Beale et al. have conducted a basic investigation on liquid water transport in both regions with an experimental-based GDL \cite{Beale_2020}, which provides a good start for current research. In this work, the geometry-based stochastic reconstruction method is used to reconstruct the GDLs with different fiber diameters and the effect of fiber diameter will be discussed.}

This paper is organized as follows: Section \ref{Section2} describes the simulation geometries we used and the stochastic reconstruction of three GDLs as well as the simulation method. Section \ref{Section3} presents the comparison results with experimental and previous results. Besides, the water transport in three geometries with different fiber diameters is qualitatively and quantitatively discussed. Finally, Section \ref{Section4} concludes the paper.

\section{Model development}\label{Section2}
\subsection{Computational domain}
{We restrict our research to low-temperature PEMFCs, with an operating temperature range of 60-80 $^{\circ}$C \cite{qu2022proton}. As shown in Fig. \ref{fig:Fig1}(a), a PEMFC device is made of a PEM in the middle and different layers in the cathode and anode regions as well, including bipolar interconnectors, GDLs, Catalyst Layers (CLs), MPLs and GCs. The basic working principle of PEMFCs is shown in Fig. \ref{fig:Fig1}(b). Hydrogen is oxidized to hydrogen ions and electrons at the anode catalyst layer. At the same time, oxygen combines with hydrogen ions and electrons through an Oxygen Reduction Reaction (ORR) at the cathode catalyst electrode layer/proton exchange membrane interface, as shown in Eq. \ref{ORR}. Then, the electric current is generated due to the continuous movement of the electrons in an external circuit. 

\begin{equation} \label{ORR}
ORR:{O_{2} + 4H^{+} + 4e{^-}\rightleftharpoons 2H{_2}O} 
\end{equation}

\begin{figure}[h]
\centering
\includegraphics[width=1\textwidth]{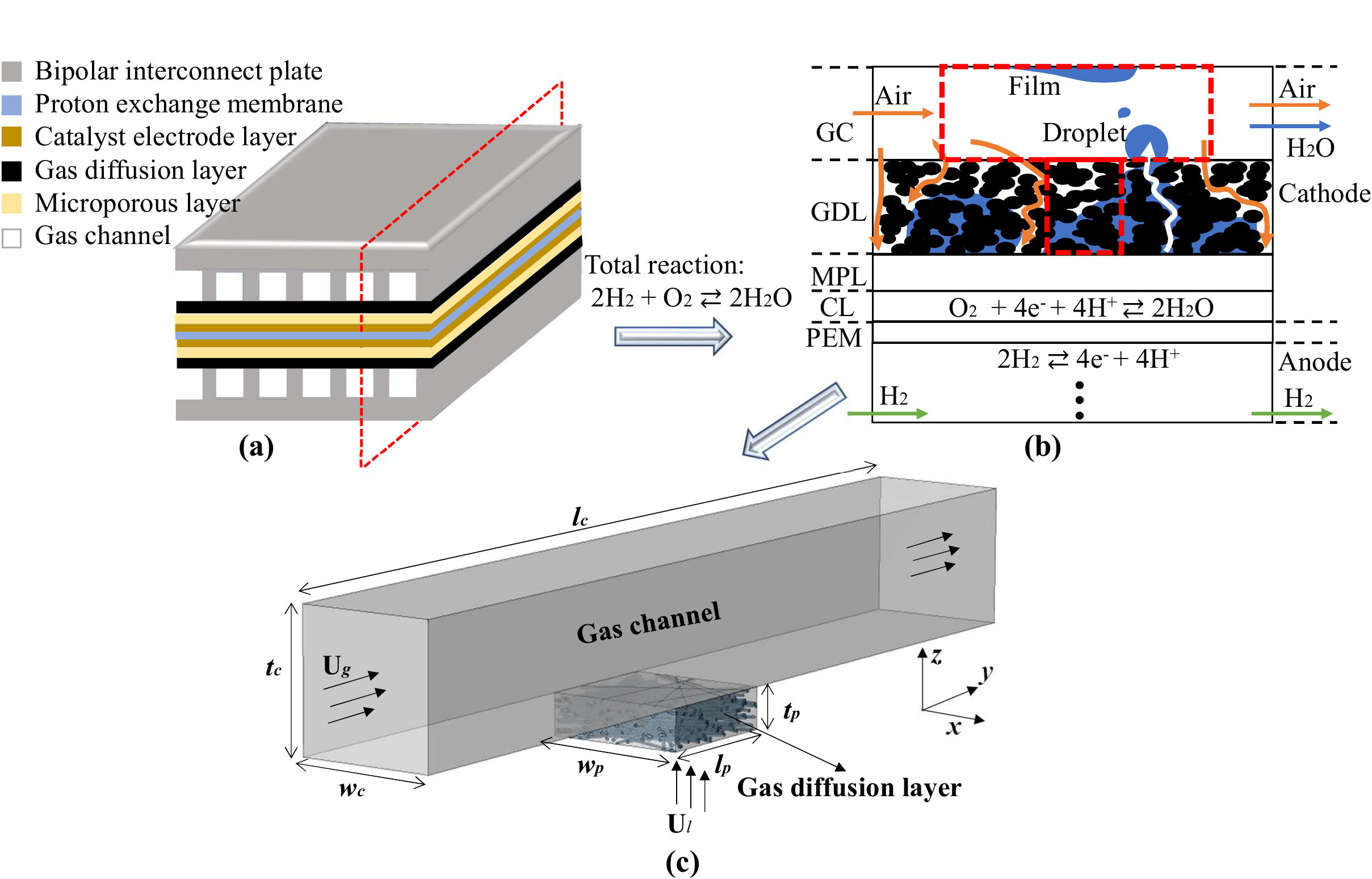}
\caption{\label{fig:Fig1} (a) Typical schematic diagram of PEMFCs; (b) Basic working principle of PEMFCs and two-phase flow in the cathode GDL and GC; (c) "T-shape" three-dimensional geometry for two-phase flow simulation in cathode GDL and GC.} 
\end{figure}

{As shown in Fig. \ref{fig:Fig1}(c), a three-dimensional "T-shape" geometry including GC and GDL is built in the Cartesian coordinate system (x, y, z). The GC length and the GDL thickness are respectively along the y-direction and z-direction. The detailed dimensions of the geometry are shown in Table \ref{tab:1}. Part of the real cathode part is selected in order to make a trade-off between computational load and research destination. The width of GDL and GC is kept the same as the real width of the GC, but the length of GDL is selected to be smaller than that of GC. Moreover, the thickness of both of them is kept with the real dimension, 1 mm and 300 $\upmu$m respectively. The input direction of each fluid phase is perpendicular to the corresponding geometric inlet, shown in Figure \ref{fig:Fig1}(c). \textbf{\textit{U$_\mathrm{g}$}} and \textbf{\textit{U$_\mathrm{l}$}} represent the gas inlet velocity and liquid inlet velocity. {\textit{l}}, {\textit{w}} and {\textit{t}} represent the length, width and thickness, respectively. The subscript {\textit{c}} and {\textit{p}} express the GC and porous GDL.}

%Considering the balance between computational load and domain size, a "T-shape" geometry, as shown in Fig. \ref{fig:Figure2}, is built to investigate the two-phase flow behavior in the combination of GC and GDL. Table \ref{tab:1} provides the dimensions of the geometry. Based on the stochastic reconstruction method, only a small part of actual GDL is generated.

\begin{table}[H]
\scriptsize
\centering
\caption{\label{tab:1} Dimensions of the reconstructed "T-shape" geometry}
\begin{tabular}{llllllllllll}
\hline
Parameters &&  Symbol  && Value     &&unit   \\\hline
Channel length  && $l_{c}$  &&7         && mm     \\
Channel width    && $w_{c}$  &&1         &&  mm     \\
Channel height    && $t_{c}$  &&1         &&  mm    \\
GDL length    && $l_{p}$  &&1         &&  mm     \\
GDL width    && $w_{p}$    &&1         &&  mm  \\
GDL thickness    && $t_{p}$  &&300         &&  $\upmu$m    \\ 
GDL fiber diameters    && $d$  &&10/15/20         &&  $\upmu$m    \\
\hline
\end{tabular}
\end{table}

\subsection{Stochastic reconstruction of GDLs} \label{ReconstructionSection}
{A GDL of the PEMFCs is usually made of carbon-fiber-based products due to their high porosity and excellent electrical conductivity. There are usually three types of GDL, i.e., carbon cloth, carbon felt, and carbon paper. In this work, the reconstructed GDLs belong to the type of carbon paper, which is made of a number of straight and finite-length cylindrical carbon fibers.  The microstructural properties of a GDL can be characterized by some parameters like fiber diameter, domain size, and porosity. To conduct the reconstruction work, the range of these parameters in some published literature is reviewed as a reference of this research, which is listed in Table~\ref{tab:widgets}. In this paper, the parameters are mainly selected based on the type of Toray TGP-H090 but have some adjustments. Besides, the bonding materials of wettability are not included in the present research, which will be considered in future work. 

\begin{table}[H]
\centering
\scriptsize
%\footnotesize
\caption{\label{tab:widgets} Microstructural physical properties of GDL material}
\begin{tabular}{lllllll}
\hline
References &Type &  Fiber diameter ($\upmu$m)  & Porosity (-) &Thickness ($\upmu$m) \\
\hline
Ashorynejad. et al. \cite{ashorynejad2019evaluation,ashorynejad2016investigation} & Toray TGP-H-120 &   10  & 0.7 &  250,300 \\ 
Mukherjee et al. \cite{mukherjee2020estimation} &SGL * &    8  & 0.75 - 0.85 &  190 ± 10 - 320 ± 10 \\
Chen et al. \cite{CHEN20168550} &Toray TGP-H-060 &7 &0.78  &103.5 \\ 
Jeon et al. \cite{jeon2015effect} &SGL 10BA &10 &0.88  &400 \\ 
Nazemian et al. \cite{nazemian2020impact}  &- &7, 8, 9 &0.75 - 0.9  &100, 200, 300\\ 
Mangal et al. \cite{mangal2015experimental}  & Toray TGP-H-090 &9.2  &0.54 ± 0.01 - 0.76 ± 0.02  & 273 ± 6 - 290 ± 4 \\
& SGL 34BA&- &0.77 ± 0.03 &260 ± 10\\
Schulz et al. \cite{schulz2007modeling} &Toray TGP-H-060 &7 &0.78  &190 \\ 
 	& SGL 10BA &7	&0.88 &380 \\
Gostick et al. \cite{gostick2006plane} &Toray TGP-H-060 &7 &0.78  &190 \\ 
 	& SGL 10BA &9.2	&0.88 &400 \\	
 	& SGL 24BA &8	&0.81 &195 \\	
 	& SGL 34BA &7.6	&0.84 &285 \\
 	& Toray TGP-H-090 &9.2	&0.8 &290 \\
 	& E-Tek Cloth ‘A’ &9.9	&0.78 &360 \\
Toetzke et al. \cite{toetzke2014three} &Freuenberg H2315 &10 &0.67 - 0.8  &154-223 \\ 
Qin et al. \cite{qiu2018electrical} &TGP-H-060  & 7–8 &0.78  &190\\
&TGP-H-090  & 7–8 &0.78  &280\\
&Freuenberg H2315  & 7–8 &0.8  &210\\
&Freuenberg H14  & 7–8 &0.8  &150\\
&Toho Tenax TCC 2660 & 10–11 &-  &260\\
&Toho Tenax TCC 3250 & 10–11 &-  &320\\
Hinebaugh et al. \cite{hinebaugh2017stochastic} &Toray TGP-H-090  & 7–11 &0.72  &263\\
\hline
\end{tabular}
\end{table}

To investigate the effect of fiber diameter on water transport, three GDLs with different fiber diameter are generated by the same geometry-based stochastic reconstruction process, which is described in \ref{appendix:reconstruction}. The porosity and dimensions of the three GDLs are kept the same, with 0.81 and 1000$\times$1000$\times$300 $\upmu$m${^3}$, respectively. One of the GDLs has a fiber diameter from the typical range of general commercial GDLs, with 10 $\upmu$m, and the other two GDLs have larger fiber diameters of 15 $\upmu$m and 20 $\upmu$m. These three GDLs with fiber diameters 10 $\upmu$m, 15 $\upmu$m, and 20 $\upmu$m correspondingly have 30, 20, and 15 carbon fiber layers. To more clearly observe the effect of fiber diameter size on gas-liquid transport in the GDL and to provide a reference for subsequent research and fabrication, the values of two of the fiber diameters are selected outside the current normal diameter range, with enough difference among them.   

Three reconstructed GDL structures and the corresponding top view snapshots are shown in Fig. \ref{fig:ReconsructionGDLs}(a-c). Compared with the SEM surface image of the Toray090 carbon paper with fiber diameter 7 $\upmu$m, shown in \cite{schulz2007modeling}, the reconstructed GDLs look very qualitatively similar in shape to the real one, especially the GDL with 10 $\upmu$m. It is clear that a smaller fiber diameter tends to lead to more numbers of carbon fibers inside the GDL, which greatly avoids the appearance of large pore space inside and even perforations. Some obvious perforations could be seen in Fig. \ref{fig:ReconsructionGDLs}(b-c). Increasing fiber diameter results in both the size and number of the perforations increasing as well.

\begin{figure}[h]
\centering
% \subfigure[]{   
% \includegraphics[width= 0.4\textwidth, trim=0 0 0 0, clip]{Paper1Reverse1Pictures/Actual GDL.png}  
% }
% \\
% \subfigure[]{   
\includegraphics[width= 1\textwidth]{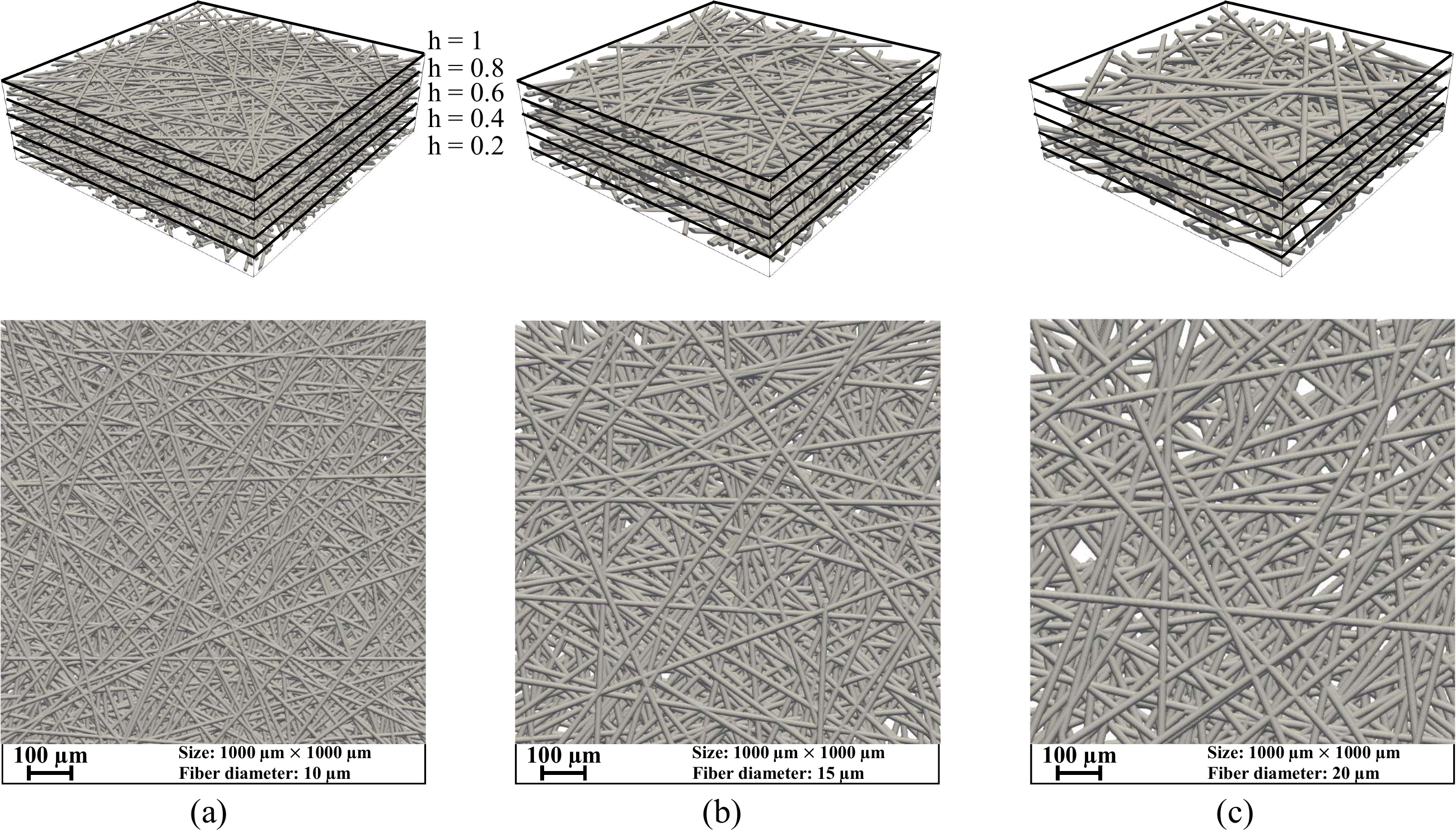}  
% }
% \subfigure[]{ 
% \centering    
% \includegraphics[width= 0.3\textwidth, trim=0 0 0 0, clip]{Paper1Reverse1Pictures/GDL2TopViewSnapshot2.png}
% }
% \subfigure[]{ 
% \centering    
% \includegraphics[width= 0.3\textwidth, trim=0 0 0 0, clip]{Paper1Reverse1Pictures/GDL3TopViewSnapshot2.png}
% }
\caption{Fibrous structures and top view snapshots of three reconstructed GDLs with different values of fiber diameter, d, with same porosity 0.81 and domain size 1000 $\times$ 1000 $\times$ 300 $\upmu$m${^3}$. ((a) d = 10 $\upmu$m; (b) d = 15 $\upmu$m; (c) d = 20 $\upmu$m.)}    
\label{fig:ReconsructionGDLs} 
\end{figure}
% (a) SEM surface image of the Toray090 carbon paper with fiber diameter 7 $\upmu$m. \cite{schulz2007modeling}. 

\subsection{Volume-of-Fluid method}
{Transient 3D simulation of liquid removal through fibrous GDL structure is conducted using the VOF method. The VOF method is a fluid fraction-based capturing technique applied to track and locate the interface of two or more immiscible fluids \cite{hirt1981volume}. This approach is based on a whole-domain formulation that treats a multi-fluid system as a single-fluid system with space-dependent physical properties. Specifically, the immiscible fluid components share a set of conservation equations. Capturing the comparable domain interface is implemented by introducing a phase volume fraction $\alpha{_i}$ of each phase in the computational cell volume, $\alpha{_i} = {V_i}/{V{_{v}}}$. Where, the volume $V_i$ is the $i$\  $(i = $gas$, $\  $ $liquid$)$ phase in the computation cell, and $V{_{v}}$ is the total volume. The total volume fraction of all fluids should be equal to 1. Both liquid water and gas exist in low-temperature PEMFCs due to the ORR reaction occurring at a low temperature. The VOF method has been widely adopted to study two-phase flow in PEMFCs \cite{CHEN20218640}. For the incompressible flow in this study (Mach number, $Ma<0.3$), the continuity equation, momentum conservation equation, and fluid-fraction advection equation in the VOF method are solved \cite{hirt1981volume}. 

Continuity equation (Mass conservation):
\begin{equation}
\nabla\cdot\mathbf{u} = 0
\end{equation}

Momentum conservation equation:
\begin{equation}\label{Momentum conservation equation}
\rho\frac{\partial{\mathbf{u}}}{{\partial t}} + \rho\nabla\cdot\left(\mathbf{u}\mathbf{u}\right) = \nabla\cdot\left[\mu\left(\nabla\mathbf{u}+\nabla\mathbf{u}^T\right)\right] -\nabla p + \rho\mathbf{g}+ {\mathbf{f}_{\sigma}}
\end{equation}

fluid-fraction advection equation:
\begin{equation}
\frac{\partial\alpha}{\partial t} + \nabla\cdot\left(\alpha\mathbf{u}\right) = 0 
\end{equation}

where, $\rho$ and $\mu$ are the density and viscosity of fluid mixture, and $\mathbf{g}$ is the gravitational acceleration. $\mathbf{f}_{\sigma}$ is the additional volumetric surface tension force. $\alpha$ is the fluid fraction, varying between 0 and 1; $\alpha$ = 0 corresponds to mesh cells that are filled with air gas, whereas $\alpha$ = 1 indicates that the cells fully contain the liquid water. Gradients of the fluid fraction are encountered only in the region of the interface between the two phases. $\mathbf{u}$ is the velocity vector, which is defined as: 
\begin{equation}
{\mathbf{u}} = \frac{\alpha\rho_{\mathrm{l}}{\mathbf{u_{\mathrm{l}}}} + (1-\alpha)\rho_{\mathrm{g}}{\mathbf{u_{\mathrm{g}}}}}{\rho}
\end{equation}

Here, $\rho_{\mathrm{l}}$ and $\rho_{\mathrm{g}}$ are the density of liquid and gas species, respectively. $\mathbf{u}_{\mathrm{l}}$ and $\mathbf{u}_{\mathrm{g}}$ are the velocity of liquid and gas phase. Moreover, the pressure, density, and viscosity of the mixture are defined as weighted averages based on $\alpha$.

The surface tension force $\mathbf{f}_{\sigma}$, in Eq. \ref{Momentum conservation equation}, refers to the continuum surface force model \cite{BRACKBILL1992335}. 
\begin{equation}
 {\mathbf{f}_{\sigma}} = \sigma\nabla\cdot(\hat{n}_c)\nabla{\alpha}
\end{equation}

Here, $\sigma$ is the surface tension of the liquid, $\hat{n}_c$ is a unit vector normal to the fluid interface, which is calculated from the fluid fraction field, 
\begin{equation}
 \hat{n}_c = (\frac{{\nabla {\alpha}}}{{|\nabla{\alpha}|}})
\end{equation}

In addition, the interaction between the fluids and the solid
surface (wall adhesion) is included as a boundary condition on the surface normal vector at the solid wall.
\begin{equation}
 \hat{n}_c = \hat{n}_s \cos{\theta} + \hat{t}_s\sin{\theta}
\end{equation}

where, $\hat{n}_s$ and $\hat{t}_s$ are the unit vectors normal and tangential to the solid boundary, respectively; ${\theta}$ is the equilibrium or static contact angle. The surface normal vector is corrected at the contact line
to satisfy the specified static contact angle boundary condition. 

In porous media, based on the Young-Laplace equation, the capillary pressure $P_c$, is defined as the difference between the wetting phase (Here is liquid water) pressure $P_l$ and non-wetting phase (Here is gas) pressure $P_g$, which depends on the average pore radius $r$, surface tension $\sigma$ and contact angle, as shown below.
\begin{equation}\label{Equation9}
 P_c = P_l - P_g = - \frac{2\sigma cos(\theta)}{r}
\end{equation}

Further information about the VOF method can be found in the previous work \cite{ANDERSSON20182961}.

\subsection{Numerical techniques and boundary conditions} \label{BoundaryConditions}
{Here, the VOF method is applied by using the interFoam solver in open-source software OpenFOAM 7.0 (OF7) \cite{jasak2009openfoam}. The aforementioned VOF equations are solved based on the PIMPLE algorithm. Considering the balance between computation efficiency and convergence rate, a simulating setting in OF7 named adjustable running time is applied with a maximum Courant number of 0.7. Besides, the discretization schemes for the convection and diffusion terms in the VOF equations are second-order accurate. The parameters including the surface tension, viscosity, and density are selected at a PEMFC operating temperature of 60 $^\circ$C, listed in Table \ref{tab:3}.}

\begin{table}[h]
\centering
\scriptsize
\caption{\label{tab:3}Species properties of two-phase flow in PEMFCs at a temperature of 60$^\circ$C}
\begin{tabular}{lllll}
\hline
Parameter & & Value ranges  & &unit   \\\hline
Water density                  & &983     & &kg/m$^3$      \\ 
Gas density                       & &1.06        &&kg/m$^3$ \\
Water viscosity             & &$4.75 \times 10^{-5}$        && m$^2$/s \\
Gas viscosity             & & $1.89 \times 10^{-5}$        && m$^2$/s \\
Surface tension             & &0.0644        && N/m\\
Average inlet gas velocity             & &(0 10 0)         && m/s     \\
Average inlet liquid water velocity          & &(0 0 0.02)      && m/s \\
Contact angle of GC top and two side walls & &45         &&$^\circ$\\
Contact angle of GC bottom wall & &150        &&$^\circ$\\
Contact angle of GDL four side walls& &150         &&$^\circ$\\
Contact angle of GDL fiber surface& &150         &&$^\circ$\\
\hline
\end{tabular}
\end{table}

{The liquid velocity at the bottom of GDL is linked with the electrochemical reaction rate. Note that the liquid inlet velocity in this study is increased by multiplying with an approximate factor of 10-35, to speed up the calculation time. A more detailed motivation for this treatment can be found in Ding et al. \cite{DING2014469}, where it is recommended to increase the inlet flow rate of liquid water by a factor of up to 1000, and the flow still presents a similar flow pattern. Ding et al. \cite{DING20107278} also concluded that the liquid flow rate has only a minor impact on the channel's two-phase flow pattern. An increased liquid flow rate yields a slightly increased channel wall water coverage ratio but has only a minor influence on the water coverage on the GDL/GC interface, which controls the detachment time and droplet size. Note that this is somewhat compensated because only a part of the channel is connected to the GDL \cite{ANDERSSON201911088}.}

{All geometric walls are assumed to be slip-free, except for two inlets and one outlet. The GC outlet is set to a boundary with a velocity boundary of pressureInletOutletVelocity, a pressure boundary of total pressure, and a zero gradient of water volume fraction.  Two inlet velocities are separately set with (0,10,0) m/s for gas and (0,0,0.02) m/s for liquid. The gas inlet $\alpha$ is defined as a uniform value of 0, and The $\alpha$ = 1 for the liquid inlet. In addition, the channel surface wettability is considered by adopting the static contact angles. Generally, a wall with a contact angle lower than 90$^\circ$ is hydrophilic. In contrast, contact angles between 90$^\circ$ and 180$^\circ$ mean hydrophobic. 
Owing to different drainage requirements, the channel walls and GDL surfaces are set with different constant (static) contact angles. For our simulation domain, the channel bottom wall and carbon fiber surface in GDL are set with a contact angle of 150$^\circ$, and the top and two side walls of the GC have a contact angle of 45$^\circ$ \cite{andersson2018modeling}.
}

{Mesh of the complex geometry is generated by applying the SnappyHexMesh in OF7, a toolbox which subtracts the inner or outer mesh region of simulation geometry from a regular block mesh that encloses the geometry, and then smooths the mesh cells based on the geometry. With the same meshing scheme and acceptable mesh qualities, the three different geometries have been meshed with about 5.4 million, 5 million, and 4.7 million mesh cells, respectively. Parallel computations on 100 central processing units are realized on a cluster. Two-phase flow dynamic simulations of three cases for 21 ms take about 260 hours.}

\section{Results and discussion}\label{Section3}
% \textcolor{red}
{A two-phase flow simulation model in PEMFCs is built based on the theoretical framework and boundary conditions we mentioned in Section \ref{Section2}. Liquid water saturation is introduced to analyze the water behavior from both qualitative and quantitative perspectives. The water saturation $S$ is defined as the ratio of water volume to empty pore volume, namely, $S = V_{water}/V_{pore}$. In Section \ref{ModelValdation}, the experimental results are used to validate the present two-phase flow study in Section \ref{ModelValdation} with different boundary conditions from that in Section \ref{BoundaryConditions}. In addition, Three simulation cases (case A, case B, and case C) are established using the three different reconstructed GDLs (See Fig. \ref{fig:ReconsructionGDLs}). Some key parameters of these cases are listed in Table \ref{tab:4}. Based on the same boundary conditions and operation parameters, two-phase flow pattern evolution in the GCs and GDLs of three cases is discussed in Section \ref{FPInGCs} and Section \ref{FPInGDLs}, and the effect of fiber diameters on the flow pattern in three GDLs is also qualitatively investigated. In Section \ref{WaterccoverGCs}, \ref{LocalSGDL} and \ref{TotalSGDLGCs}, some quantitative indexes are also adopted to analyze the effect of fiber diameters, i.e. water coverage ratio in GCs as well as local and total water saturation in GDLs. Furthermore, it is necessary to mention that the summarized maximum relative error in this study is less than 1.5 $\%$, which is in an acceptable range.}
% The distribution of water in the middle slice of GDL at different times, the evolution of water in the whole domain over time, and the change of water saturation at different slice locations are all discussed.

To characterize the water behavior, some parameters such as the total water saturation $S_t$, local water saturation $S_l$, and porosity $Por_l$ in this research are introduced by morphological processing and necessary pixel analysis of a series of in-plane GDL slices. All the slices have a very thin and uniform distance of 1 $\upmu$m. Thus, based on the idea of mathematical differentiation, the liquid area $A_{l,water}$ and pore area $A_{l,pore}$ in each slice are used to calculate the water saturation and porosity, as shown in Eq. \eqref{33} - Eq. \eqref{34}. $A_{l,total}$ is the total pixel area of each slice.}
\begin{equation}
    S_t = \frac{V_{t,water}}{V_{t,pore}} 
    \label{totalSEq}
\end{equation}
\begin{equation}
    S_l = \frac{V_{l,water}}{V_{l,pore}} = \frac{\int_{z}^{z+ \Delta z} A_{l,water}dz}{\int_{z}^{z+ \Delta z} A_{l,pore}dz} \approx \left|\frac{A_{l,water}\cdot\Delta z}{A_{l,pore}\cdot\Delta z}\right|_{(\Delta z \to 0)} = \frac{A_{l,water}}{A_{l,pore}} 
    \label{33}
\end{equation}
\begin{equation}
    Por_l = \frac{A_{l,pore}}{A_{l,total}} 
    \label{34}
\end{equation}

\subsection{Model validation}\label{ModelValdation}
% \textcolor{red}
{To prove the effectiveness of the methods we used, the local liquid water saturation of the reconstructed GDLs in the through-plane direction is validated with the X-ray tomographic experiment data in \cite{fluckiger2011investigation} and compared with previous study results \cite{niu2019two}. 
% Similar to \cite{niu2019two}, the same constant pressure difference boundary condition between the GDL top and bottom surfaces is also used here, with 1000 Pa. This value is smaller than the liquid breakthrough capillary pressure (about 5000 Pa \cite{niu2019two}). Therefore, the liquid water just can arrive at a finite height of the GDL. 
% Thus, the liquid water saturation in Fig. \ref{fig:Valiat1} is different from that in Fig. \ref{fig:Valiat2}. 
To have a rigorous comparison, the GDL structure case with 10 $\upmu$m fiber diameter is selected to simulate alone without the GC. Most of the conditions are set to be the same as the literature \cite{niu2019two}, which are close to the experimental conditions. The contact angle of carbon fiber is $\theta$ = 109 $^\circ$, and a pressure difference $\Delta p$ = 1000 Pa is applied between the GDL inlet (bottom side) and outlet (top side). The fiber diameter and porosity of the experimental GDL are 10 $\upmu$m and 0.78, which are close to the value of this study, and the referring literature \cite{niu2019two} has the value of 8 $\upmu$m and 0.73. In this validation case, the pressure difference drives the water upward, while the capillary force of the porous GDL hinders the water from rising because of the setting of the hydrophobic boundary condition of the carbon fibers. In essence, the pressure difference of 1000 Pa is smaller than the liquid breakthrough capillary pressure (about 5000 Pa \cite{niu2019two}). Therefore, the liquid water stops rising at a finite height lower than the GDL thickness once the two forces reach equilibrium. According to Fig. \ref{fig:Valiat1}, qualitatively speaking, the simulated local water saturation has a similar trend with both compared results, showing a fast decrease of water saturation nearby the GDL inlet and reaching 0 at one relative thickness position. Specifically, the results are closer to that in \cite{niu2019two} and both of them are lower than the experimental results. The liquid water stops rising at a relative thickness of around 0.2 for both studies. However, the experimental case has a larger water saturation and higher stop-lift position. There are some reasons that can cause the difference between this study with the compared studies. Firstly, the contact angle of real GDL fibers is not evenly distributed. Besides, the porosity distribution of GDL in the numerical and experimental studies is not the same. Finally, other water resources may be present in the experimental GDLs due to incomplete drainage in the last experiment.}

\begin{figure}[H]
\centering
\includegraphics[width=0.6\textwidth]{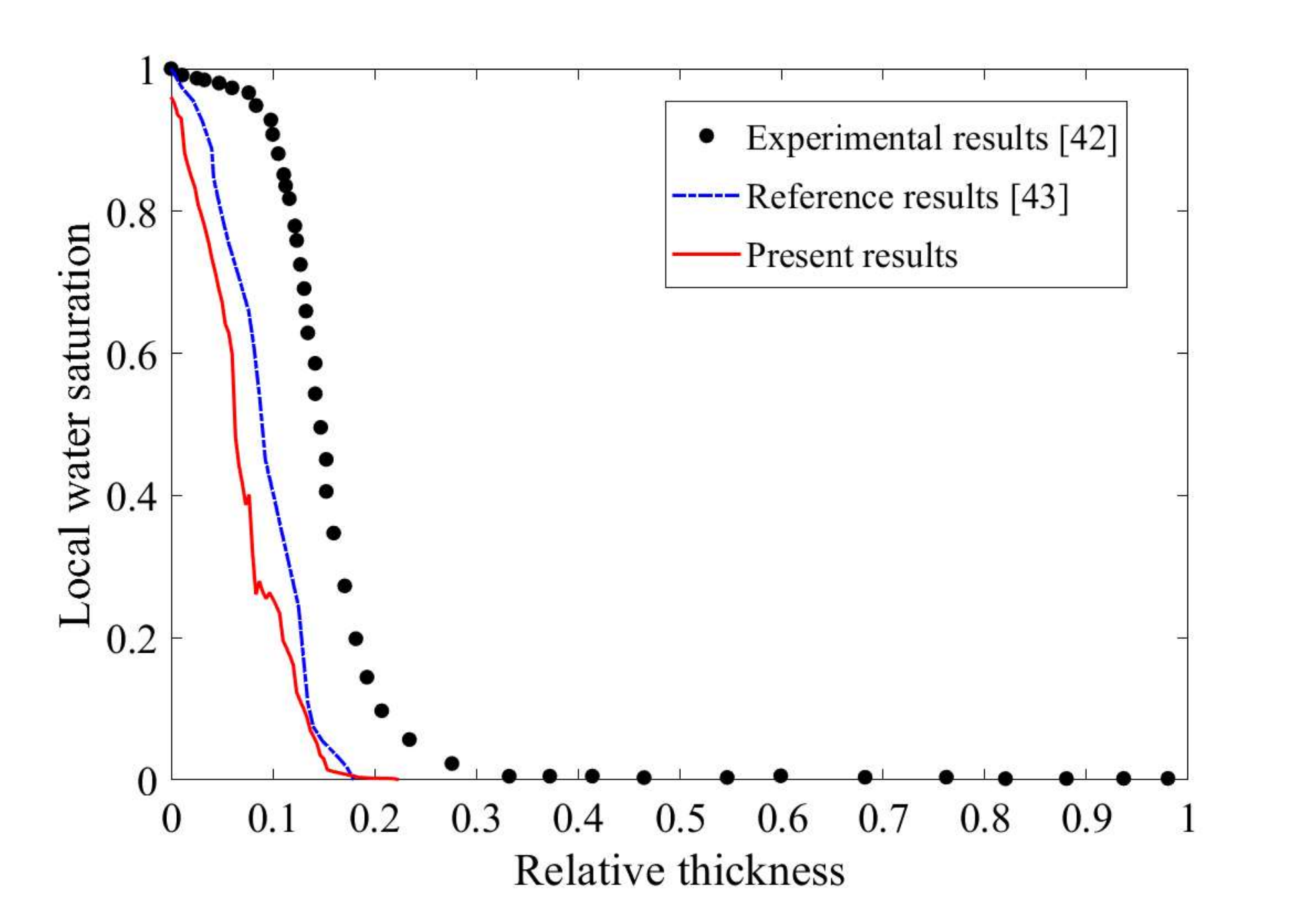}
\caption{\label{fig:Valiat1} Comparison of the local liquid water saturation in the through-plane direction of GDLs between the present results and X-ray tomographic experiment data \cite{fluckiger2011investigation} as well as previous study results \cite{niu2019two}. Operation conditions: contact angle of carbon fiber is $\theta$ = 109 $^\circ$, and a pressure difference $\Delta p$ = 1000 Pa is applied between the GDL inlet and outlet side.}
\end{figure}

Due to the lack of experiment or simulation studies that focus on two-phase flow in both GDL and GC simultaneously, the liquid behavior in the GC is qualitatively similar to the previous study of Andersson et al. \cite{andersson2018modeling} that has done the experiment and simulation to observe the droplet transport in a small GC. Some important liquid water evolution in the GC can be observed in both experiment and simulation, for example, the water droplet formation, attachment, and detachment at the GDL/GC interface; water film or slug generation in channel side walls; and transport on the channel top wall. More details can be found in the previous study \cite{andersson2018modeling}. 

\subsection{Two-phase flow pattern evolution}\label{FPInGCsAndGDLs}
\subsubsection{Two-phase flow pattern evolution in GC}\label{FPInGCs}
Different from the validation section, the boundary conditions mentioned in Section \ref{BoundaryConditions} are used in the following sections. With a fixed liquid flow rate boundary condition, the liquid break through the GDL. In this section, we analyze the general two-phase flow pattern evolution in the GC based on 3D contour plot results of three cases at 6 time steps (6 ms, 9 ms, 13 ms, 15 ms, 19 ms, and 21 ms), shown in Fig. \ref{fig:3Dwater3Cases}. The results indicate several stages: droplets breakthrough at the GDL/GC interface as well as accumulation, combination, attachment, and detachment in the GC regions. Liquid water tends to break through in the form of small droplets, gradually growing up as the liquid continues to rise. Owing to the wettability difference between the GDL/GC interface and the other three walls, the droplets close to the channel side walls always attach the side walls in a flat shape. Moreover, the droplets in the top center region of GDL keep an irregular spherical shape. However, all the liquid in the GC is affected by air gas flow. Therefore, the spherical droplets on the top surface of the GDL detach instead of accumulating continuously at the same location. The separated droplets then continue to move forward along the Y+ direction and randomly merge with other droplets. During this period, a large spherical droplet can be formed by droplet combination, as can be seen, for example, from the results for case A from t = 13 ms to t = 15 ms. On the other hand, the flat water droplets on the hydrophilic sidewalls are more likely to aggregate into large water droplets or even large liquid water clusters, see the results for the three cases after t = 9 ms.

The stages of accumulation, combination, attachment, and detachment in all three GCs occur randomly. To gain a clearer understanding of the reasons, we analyze the forces on the droplets and liquid clusters attached to the sidewalls. Considering the general formula for wind load, which is $F = A \times P \times C_d$, where $A$ is the force area of the object, $P$ is the wind pressure and $C_d$ is the drag coefficient. The force coming from the channel airflow is not correctly determined by the size of droplets and liquid clusters but by effective contact area $A_{gl}$ of liquid and air which is perpendicular to the direction of airflow. Besides, the “adsorption force” of walls on liquid has a positive correlation with the effective contact area $A_{sl}$ of liquid and walls. Moreover, we introduce the Bond number, $Bo = \Delta \rho g R^2/\sigma$ to assess the effect of gravity force. Even we assume $R$ to be large enough with a value of 300 $\mu$m considering the GDL average pore size and thickness. The estimated $Bo = $0.014 is far below 1, which means the gravity force makes less impact. Therefore, it is clear to explain that the small droplets on the front side wall of the channel in case A hardly move (see results of case A after t = 13 ms), and big droplets or liquid clusters on the side walls flow out slowly. Furthermore, once the droplets and slugs on the side walls reach the top wall, the liquid water quickly spreads on the top wall. In addition, under the action of the "adsorption force" of the top wall to the membrane and the faster flow rate, the water clusters tend to break up and separate from the side walls, thereby forming a thin liquid film on the top wall. As this liquid film moves rapidly outward, some of the water on the sidewalls and channels combines and is then dragged away (see case A, from t = 15 ms to t = 19 ms).

At the GDL/GC interface of case A, the droplet size and total droplet volume are smaller than that of the other two cases, but there are more locations for droplets to break through. Specifically, at t = 6 ms, several droplets in case C break through the GDL/GC interface, and case B has a smaller number of droplets. However, for case A, the liquid water is still concentrated within the GDL. The results show that the drag of the carbon fiber microstructure can significantly affect the water movement even if the fibers are defined as hydrophobic. Furthermore, the droplets of different sizes continuously break through the GDL/GC interface for all three cases between 9 ms and 21 ms. Case B and case A have less water in the GC in comparison to case C, and the flow of liquid water in the GC of case A is more intermittent. Moreover, the water movement in the GC becomes more continuous as the fiber diameter increases. The results indicate the water in case C flows out of the GDL faster because the larger pore size reduces the resistance of the fibers and simplifies the liquid flow path. 

\begin{figure}[H]
\centering
\includegraphics[width=0.95\textwidth]{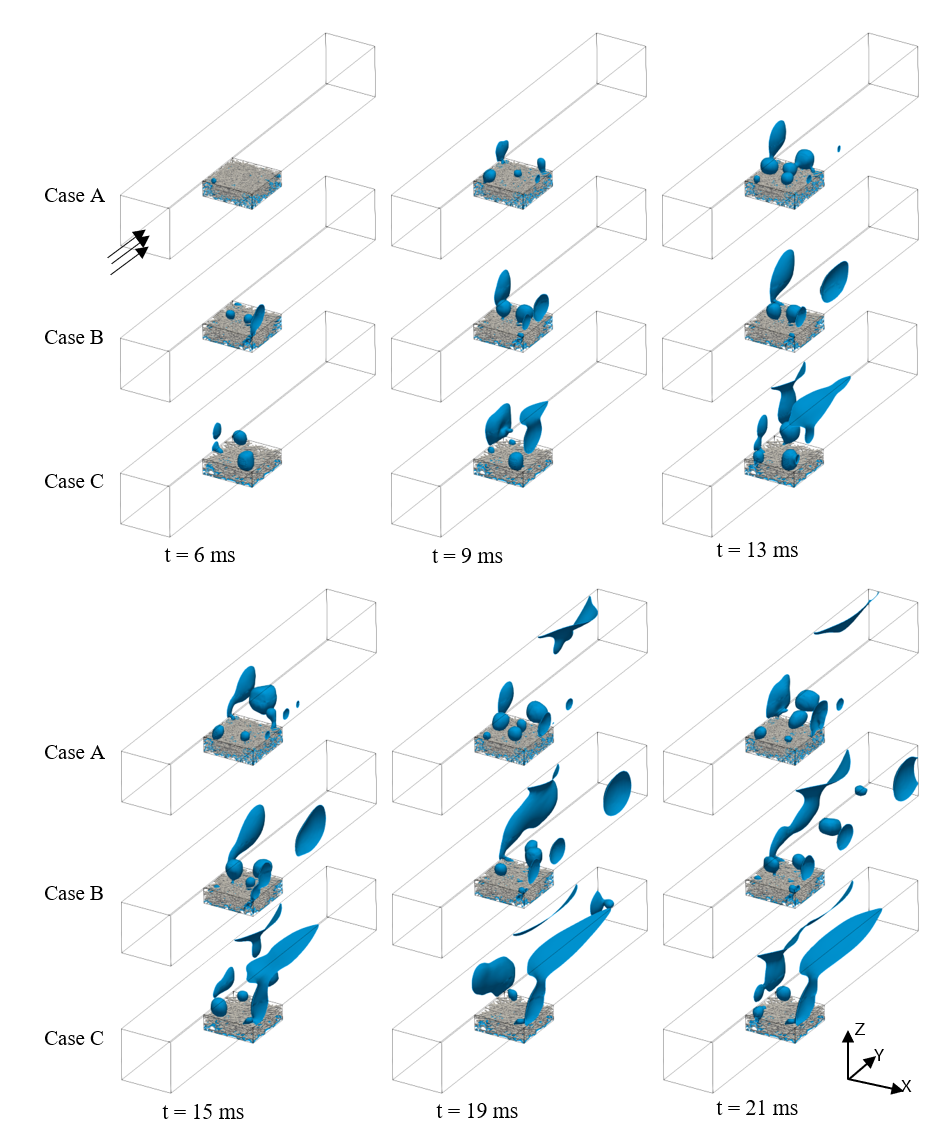}
\caption{\label{fig:3Dwater3Cases}Comparison of time-varying water transportation in the GC for three cases at t = 6 ms, 9 ms, 13 ms, 15 ms, 19 ms, and 21 ms with 3D contour plots. As time increases, case A has the slowest droplet detachment in the GDL/GC interface. case A also generates the smallest droplets in the GC, which make water removal out of the channel easier, resulting in fewer amounts of water that accumulates in the GC compared to the other two cases.}
\end{figure}
% The fastest water removal is observed in case C.
% Combining the results of the figure \ref{fig:2DwaterComparison}, 
% we conclude that the water in case C flows out of the GDL faster because the larger pores reduce the resistance of the fibers and simplify the liquid flow path. However, the obvious shortcoming is the excessive accumulation of droplets in the GC. Reducing the fiber diameter favors an increase in the number of smaller pores, thereby mitigating the accumulation of droplets in the GC due to the prolonged rise time of GDL droplets. 

\subsubsection{Two-phase flow pattern evolution in GDL}\label{FPInGDLs}
To get more details from water intrusion to droplet breakthrough in the GDL, the time-varying cross-section water distribution at three GDL relative thicknesses (h = 0.2, 0.6, and 1) is shown in Fig. \ref{fig:2DwaterComparison}. These 2D horizontal slices are taken along the X-Y section and correspond to an actual thickness of 60 $\upmu$m for the relative thickness h = 0.2. The water phase in the GDLs goes through several stages: water intrusion, water diffusion and void filling in the horizontal direction, water rise in the vertical direction, and water breakthrough. The increase of the water distribution area after the water reaches a certain slice means the diffusion and filling behavior. At the same time, the water keeps rising through each layer before breaking through the interface. 

From the water removal process, it is found that the water distribution at low relative positions (h=0.2, 0.6) tends to stabilize after a certain time. The stabilization of water distribution means that the droplet and cluster size as well as the liquid appearance position all have little change.
Due to the water breakthrough at h=1, the shape, size, and positions of the water distribution change greatly, showing the instability characteristics. The reason is that once liquid water reaches the GDL/GC interface, the water breaks through the GDL and creates droplets. At the same time, they are affected by the gas flow in the GC along the Y+ direction. Therefore, the deformation of the droplet during the breakthrough process goes through several stages: droplet growth, detachment, combination, and forced motion. Due to the randomness of droplet behavior such as separation and combination, the droplet flow is unstable.  

\begin{figure}[H]
\centering
\includegraphics[width=0.95\textwidth]{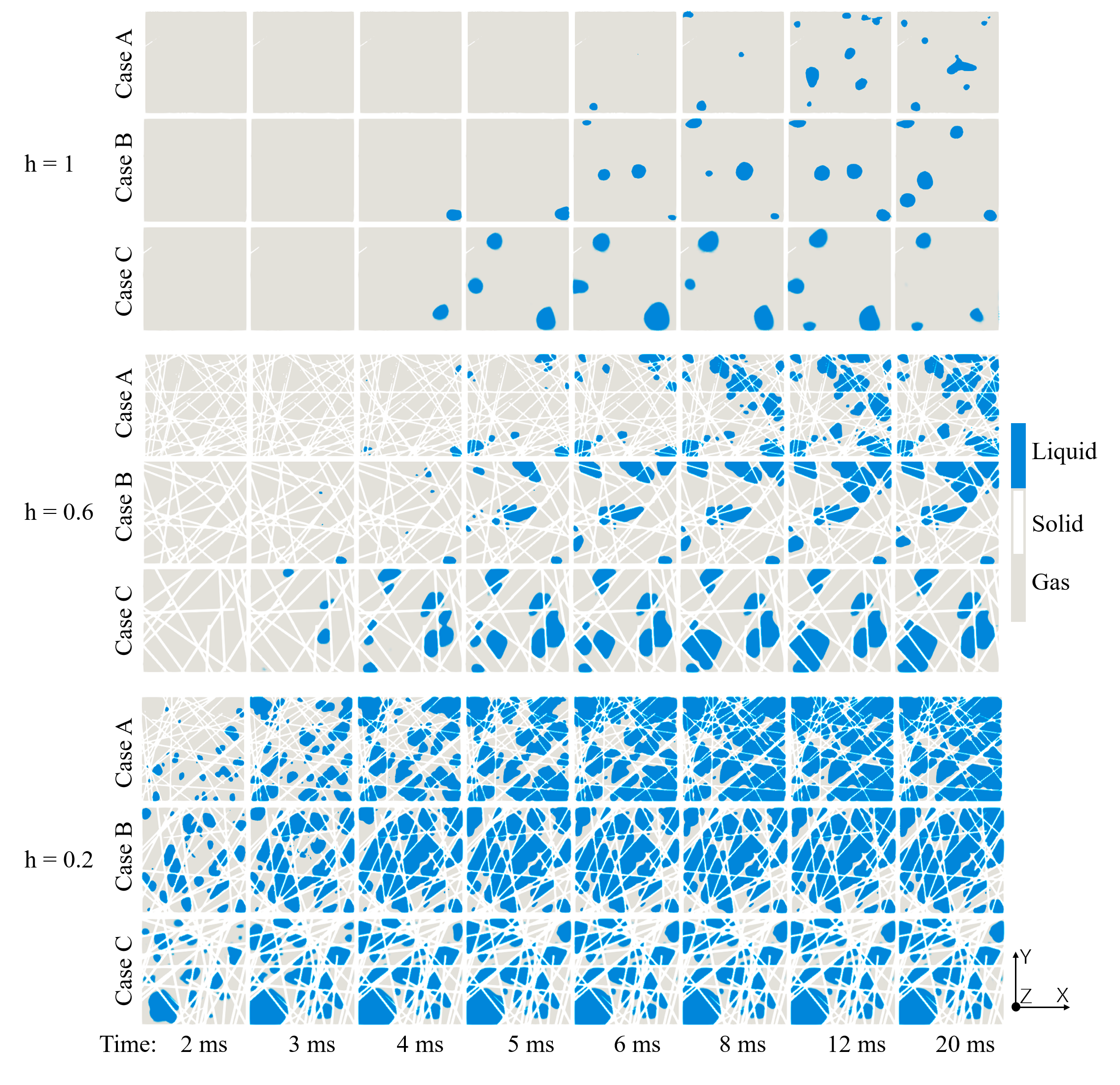}
\caption{\label{fig:2DwaterComparison}Time-varying water distribution at five relative thickness h of extracted from X-Y cross-sections of the GDL. The blue color and gray color respectively represent the liquid water and air gas phases. The  two-phase flow is a typical capillary fingering pattern that reaches a steady state after the water breakthrough. It includes several main stages: water invasion, water spread and void filling, water rising, and droplet breakthrough. Stability below h = 1 and instability at h = 1 are observed. Case A has a larger water distribution area and smaller water clusters compared with the other two cases. Besides, case A takes a longer time to reach a steady state.}
\end{figure}

In addition, the water breakthrough time is related to some drag forces such as surface tension force, viscous force, gravity force, and resistance of complex geometry structures. To analyze the effect of these forces, three independent dimensionless parameters are chosen to describe the system, which is the capillary number $Ca$, viscosity ratio $M$, and $Bo$. $Ca$ and $M$ are respectively defined by the following equations: $ Ca = {\mu_l V_l}/{\sigma}$, and $M = {\mu_l}/{\mu_g}$. Where, the subscript $l$ and $g$ represent inlet liquid and the gas in GDLs, respectively. $\mu_l$ is the dynamic viscosity, and $V_l$ is the Darcy velocity of the invading fluid. $\Delta \rho$ is the density difference and $R$ is the water cluster radius. As we discussed in Section \ref{FPInGCs}, the gravity force can be ignored. Besides, Lenormand et al. \cite{lenormand1988numerical} developed a phase diagram to map the flow patterns in the displacement patterns in log$_{10}$M-log$_{10}$Ca plane, including the viscous fingering (characterized by fingering propagation toward the outlet), capillary fingering (characterized by void-filling in the transverse/backward directions), stable displacement and the crossover zones. In this research, log$_{10}$Ca and log$_{10}$M are calculated as -9.1 and 3.2, which is well within the regime of capillary fingering, indicating that capillary force is more important than viscous force. Therefore, capillary force is the dominant force. In addition, under the continuous overcoming of the capillary force, the kinetic energy of the fluid is consumed. The ratio of the liquid water area to the pore area of the entire slice is the largest near the GDL water inlet and decreases rapidly with the increase of h.

Compared with case B and case C, case A has the same porosity and domain size but a smaller fiber diameter, resulting in an increase in the number of fibers in the GDL structure and a decrease in average pore size. Moreover, a small average pore size leads to a large capillary force in the condition of hydrophobic fiber walls, thereby hindering liquid water movement toward the GC. Thus, the liquid water distribution in case A is more complex, compared with the other two cases, showing smaller but with an increased amount of water clusters at each relative position, especially the part close to the water inlet. Additionally, a smaller pore size gives an increased flow resistance because of more complex flow patterns. Therefore, case A tends to need a longer time to have water distribution appearance in each position as well as to reach an almost stable flow. For example, in case C, it takes 3 ms to be stable at h = 0.2, but it takes almost 6 ms in case A. At t = 3 ms, new water distribution appears in position h = 0.6, while at the same position, no water appears in case A. Besides, at h = 1, From case A to case C, time-varying water distribution has a larger water cluster size and less number, indicating larger droplet size and fewer droplet breakthrough positions.

% To optimize the water removal, the following strategies could be considered after the analysis of water transport in GDL and GC, i.e., adjusting the range of GDL fiber diameters, changing the wettability of both the carbon fibers and the GC walls, controlling the air flow rate in the GC and altering the geometric size of the GDL as well as GC. 

\subsection{Water coverage ratio in GCs}\label{WaterccoverGCs}
We present quantitative results by adopting several important parameters such as the water coverage ratio of GC walls, local water saturation in GDLs as well as total water saturation in GDLs and GCs. To some extent, they realize a further verification of the aforementioned qualitative analysis results. An image-processing method is adopted to extract the water coverage ratio and local water saturation at different positions. Besides, the total water saturation is obtained by calculating the ratio of total liquid volume to the volume of a specific space, based on a post-processing function in Paraview software.  

The water coverage ratio is defined as the ratio of the contact area between the liquid and a surface to a certain area of the surface, used to identify dynamic water behavior in GCs of a PEMFC \cite{zhang2022numerical}. In the cathode part of a PEMFC, the injected reduction reactants start to diffuse from the GDL/GC surface, therefore, the water coverage ratio of this surface is a key parameter that indicates the effects of liquid water on gas diffusion. The water coverage ratio of the channel walls and the GDL/GC interface indicates the amount and distribution of water in a channel. Generally, a lower surface water coverage ratio in the channel region tends to bring the fuel cell a better performance. 

\begin{figure}[H]
\centering
\subfigure[]{\includegraphics[width=0.49\textwidth]{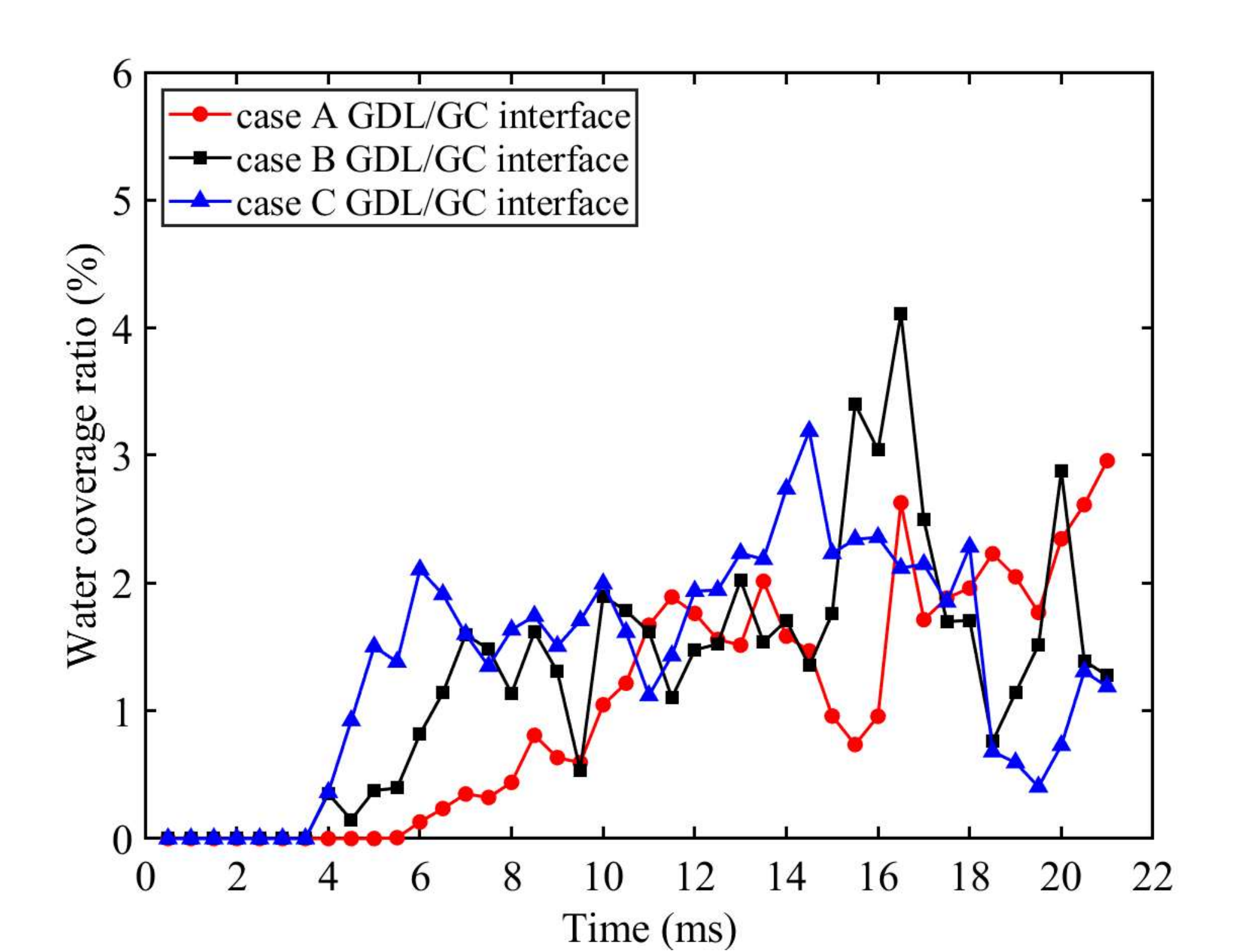}}
\centering
\subfigure[]{\includegraphics[width=0.49\textwidth]{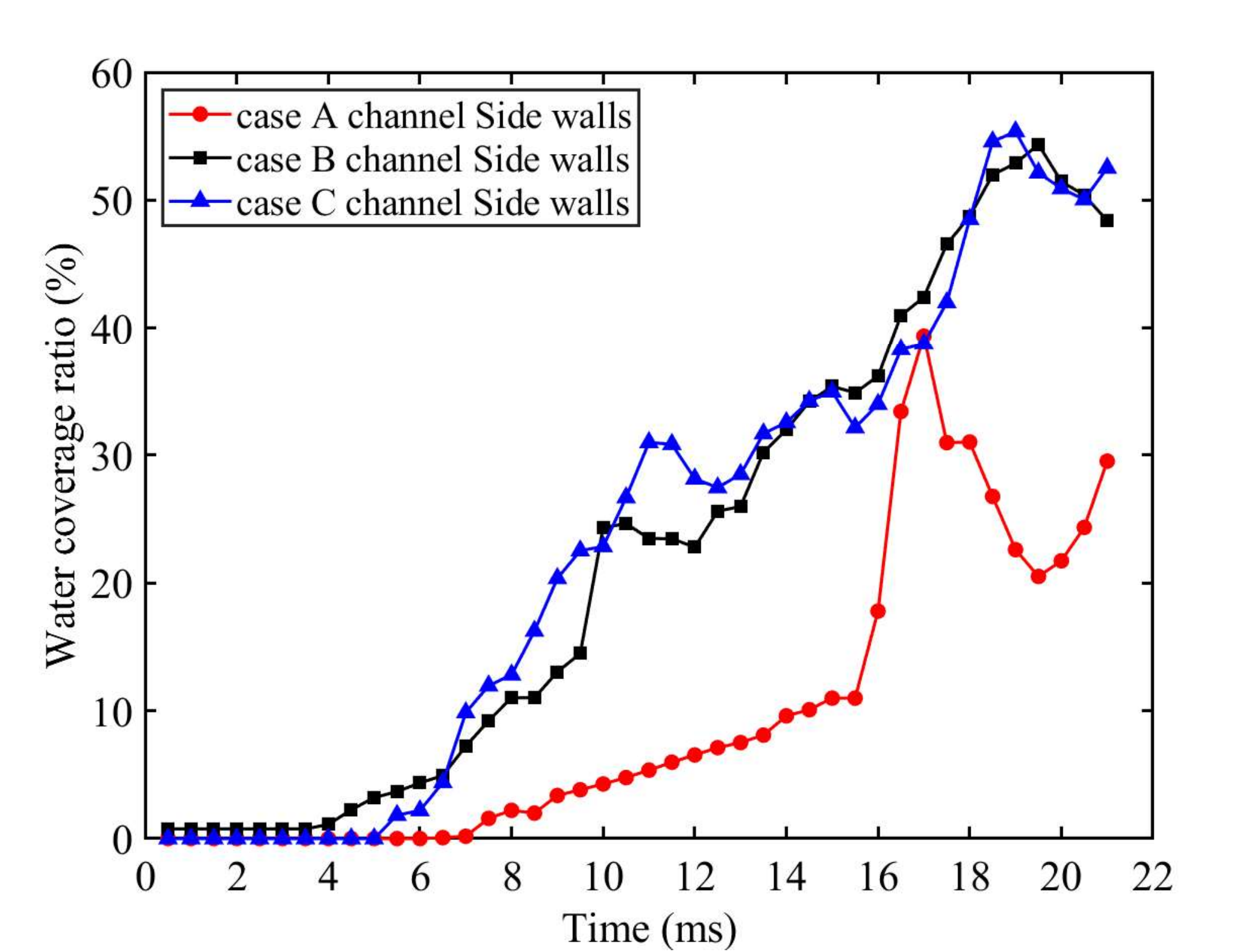}}

\centering
\subfigure[]{\includegraphics[width=0.49\textwidth]{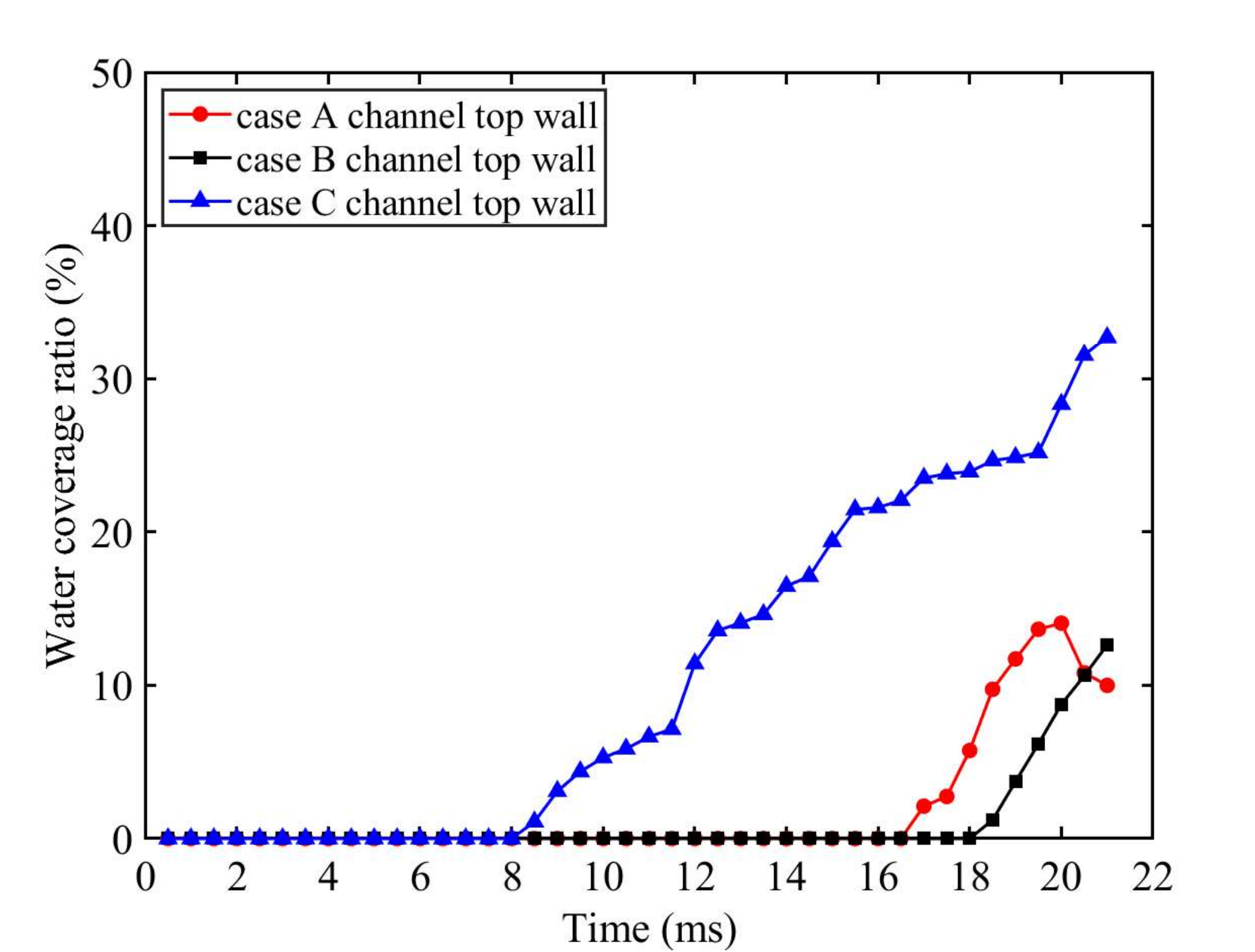}}
% \subfigure[]{\includegraphics[width=0.49\textwidth, trim=0 180 10 210,clip]{Paper1Reverse1Pictures/AverageCoverZhuzi.pdf}}
\caption{\label{fig:Watercoverageratio} Time-varying water coverage ratio of three sets of surfaces in three cases. (a) GDL/GC interface. (b) Channel side walls. (c) Channel top wall. The water coverage ratio at the GDL/GC interface oscillates irregularly, while the water coverage on other surfaces tends to increase after a certain time. }
\end{figure}

The time-varying water coverage ratio of the two side walls and the top wall of a GC as well as the GDL/GC interface is investigated, as shown in Fig. \ref{fig:Watercoverageratio} (a)-(c). Notes that only the 4/7 region close to the GC outlet is used for the ratio calculation because almost no water appears in the other GC region. Due to the different wettability of the GC walls, the water coverage ratio of the side walls and top wall is higher than that of the GDL/GC interface, referring to Fig. \ref{fig:3Dwater3Cases}. Fig. \ref{fig:Watercoverageratio}(a) shows the water coverage ratio of the GDL/GC interface. The results also indicate droplet breakthrough occurs between 3.5 ms and 4 ms for both case B and case C, and between 5.5 ms and 6 ms for case A. Besides, the water coverage ratio of the GDL/GC interface fluctuates greatly compared with that of the other channel walls. The rise in the fluctuations might be caused by the growth, accumulation, and merging of droplets at the GDL/GC surface, whereas the fall means the detachment and break-up of the droplets. Compared to the results in the work of Zhang et al. \cite{zhang2022numerical} which mainly focuses on two-phase flow in GCs, the fluctuation of the water coverage ratio in this research is stronger and irregular. In their study, a plane with circular holes of the same size and regular distribution is regarded as the interface of GDL/GC in \cite{zhang2022numerical}, which makes the water breakthrough more stable and water coverage oscillation more regular. In addition, more water coverage fluctuation states were obtained in their work, which is not within the scope of this study. We are concerned with the two-phase flow behavior before flowing out of the GCs.

The liquid water detaching from the GDL/GC surface first passes through the side walls, then reaches the top wall. In Fig. \ref{fig:Watercoverageratio}(b), the water coverage ratio of the channel side walls shows an increasing trend, where case B and case C are close to each other, and both of them have a higher water coverage ratio than case A. In case A, a large jump in the water coverage ratio of side walls happens because a big droplet merges with water on side walls after t = 15 ms, referring to the results at t = 15 ms in Fig. \ref{fig:3Dwater3Cases}. In Fig. \ref{fig:Watercoverageratio}(c), water coverage of all three top walls is observed at a later time compared with that of other surfaces. In this regard, the top wall of case B has an early water appearance and a higher water coverage ratio than the other two cases.
In 21 ms, each case has the lowest water coverage ratio at the GDL/GC interface compared with that on the other surfaces, lower than 5 $\%$. Case C has the largest water coverage on the three sets of surfaces, which means the largest amount of water accumulated in the GC. In addition, in comparison to case B and case C, case A almost has a low water coverage ratio on the side walls and bottom wall while the value on the top wall is in the middle, expected to be a lower water accumulation in the GC.

% Fig. \ref{fig:Watercoverageratio}(d) shows the time-averaged water coverage ratio of GC walls. It is found that case C has the largest average water coverage on the three sets of channel walls, with 23.35 $\%$ on the side walls, 1.39 $\%$ on the bottom wall, and 10.56 $\%$ on the top wall. In addition, case A has the lowest value on the side walls and bottom wall while the value on the top wall is in the mediate. The average water coverage ratio on the side walls of case A is less than half of the other two cases, with 9.59 $\%$. Each case has the lowest ratio on the bottom wall compared with that on the other walls, with 1.05 $\%$, 1.29 $\%$, and 1.39 $\%$ for case A, case B, and case C, respectively. In general, the water coverage ratio can reflect the accumulated water in GCs. Therefore, the lower ratio of case A could bring better operation conditions for the fuel cell compared with the other two cases.

\subsection{Local water saturation in GDLs}\label{LocalSGDL}
Fig. \ref{fig:GDLLocalS} shows the time-varying local water saturation of three cases. As Fig. \ref{fig:GDLLocalS} (a) shows, five cross-section positions (at relative thickness h = 0.2, 0.4, 0.6, 0.8, and 1) of the GDLs are selected to extract the local water saturation based on the volume fraction distribution. The results of case A, case B, and case C correspond to Fig. \ref{fig:GDLLocalS} (a-c). The local water saturation in each position of the three cases shows a rising trend at first and becomes almost stable afterward. It is found that the local water saturation gradually decreases from h = 0.2 to h = 1 of the GDLs, which also could be seen in Fig. \ref{fig:ReconsructionGDLs} (a). In addition, the local water saturation for case A from h = 0.2 to h = 0.8 is larger than those at the same locations in the other cases, while at h = 1, case A has the smallest local water saturation, and case C has the largest one. Furthermore, from case A to case C, the local water saturation difference between h = 0.8 and h = 1 becomes smaller, especially since it is almost zero in case C. Moreover, the saturation at h = 0.4 and h = 0.6 is also close. Here, we didn't get a clear reason, but the pore size distribution may account for the difference, which will be explored in future work.

The vertical black dash and solid lines in each figure represent the water breakthrough time and the time when liquid water begins to stabilize, respectively. It is apparent that the stabilization moments are a little slower than the breakthrough moments in each case. Moreover, the time difference between these two water actions gets smaller and smaller from case A to case C. The time difference is greatly related to the liquid flow velocity in the GDLs, which indicates less drag force for case C. From h = 0.2 to h = 1, the beginning of stabilization in the cross-section position requires a longer time. It is found that case A takes a long time to stabilize at every position of h than the other two cases, and case C stabilizes almost simultaneously at all positions of h. However, the stabilization of local water saturation in case A has less oscillation compared with those of case B and case C.

\begin{figure}[H]
% \centering
% \subfigure[]{\includegraphics[width=0.49\textwidth, trim=10 60 160 110,clip]{Paper1Reverse1Pictures/GDLP.pdf}}
\centering
\subfigure[]{\includegraphics[width=0.49\textwidth]{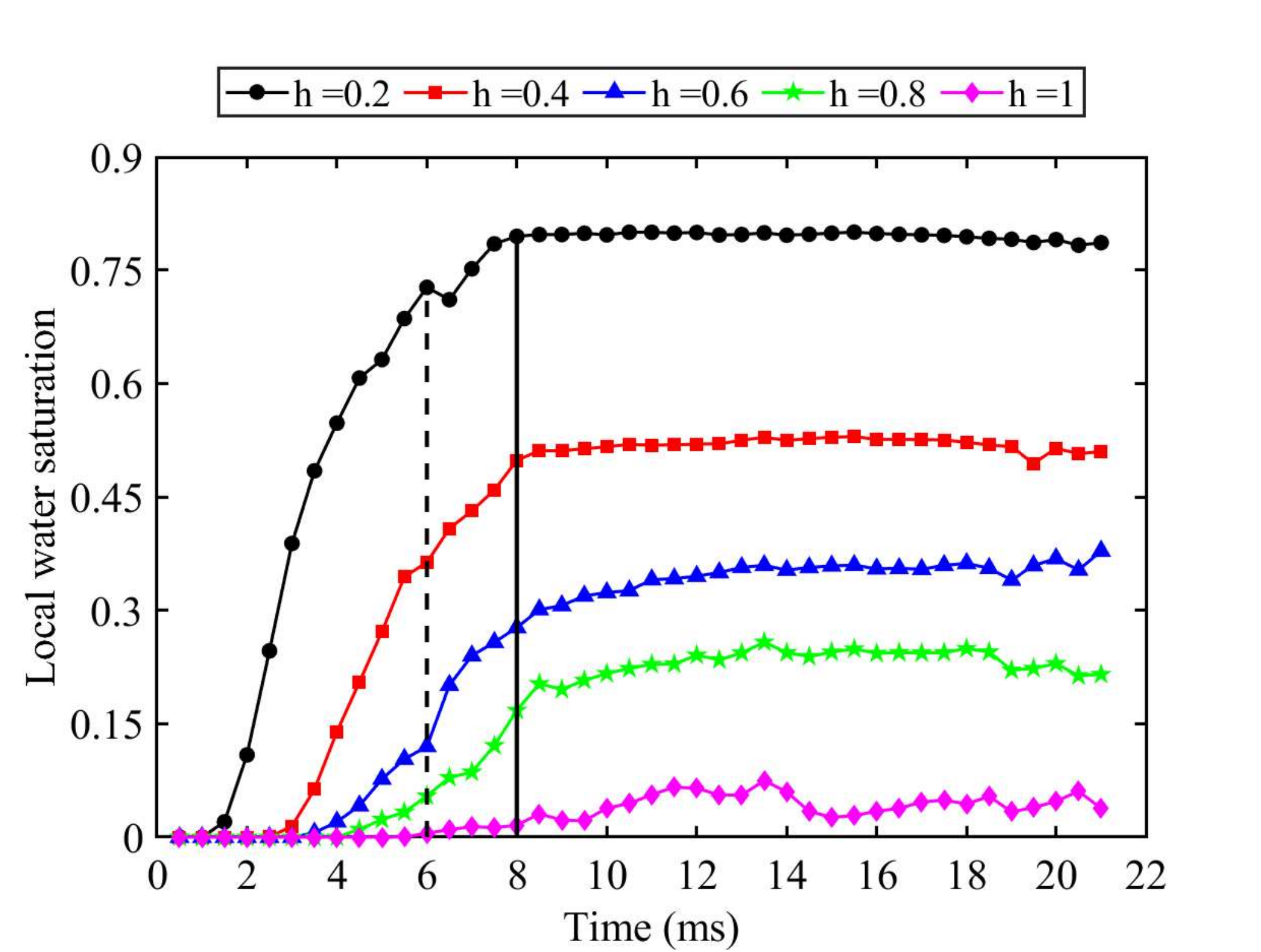}}
\centering
\subfigure[]{\includegraphics[width=0.49\textwidth]{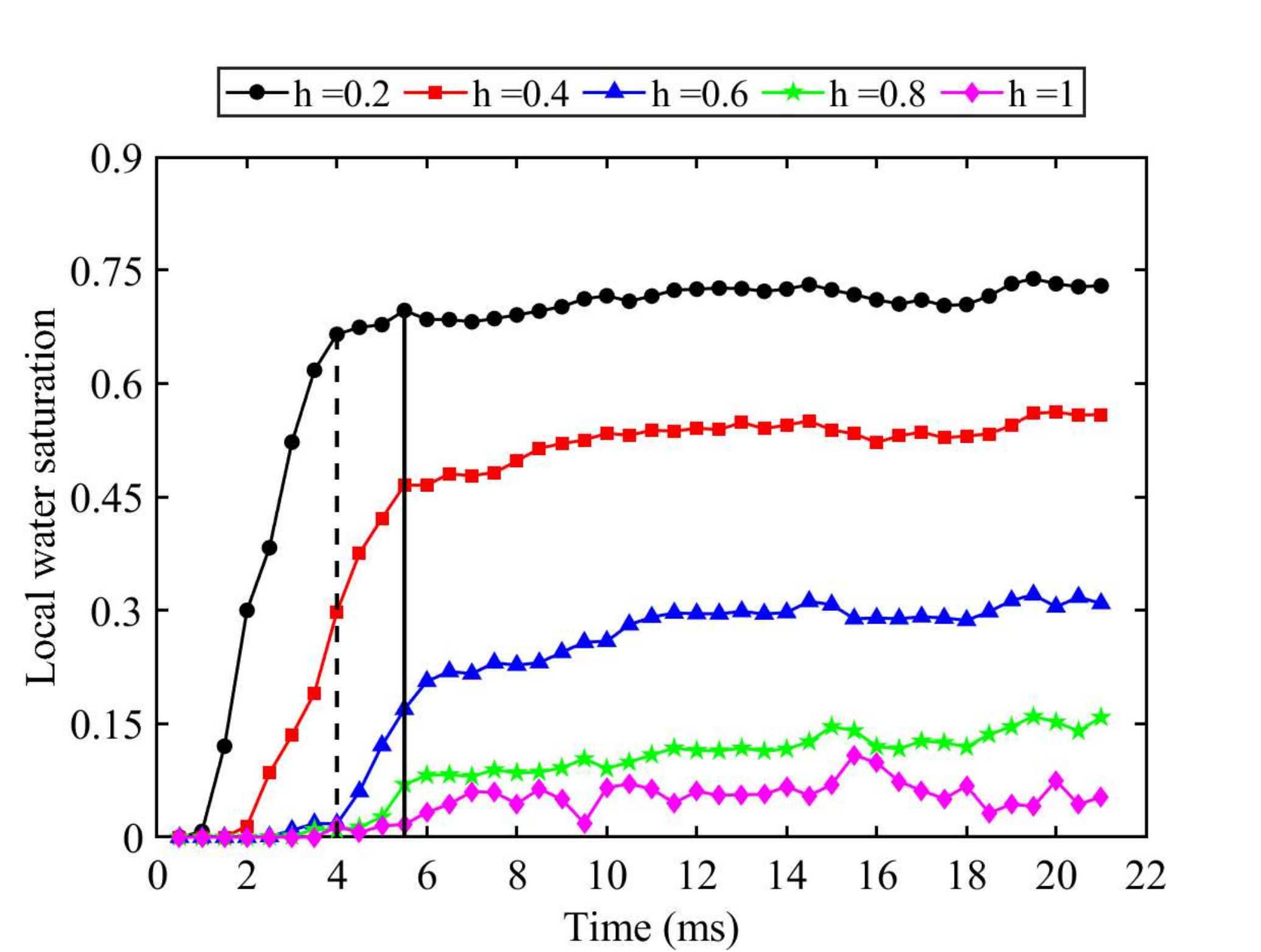}}

\centering
\subfigure[]{\includegraphics[width=0.49\textwidth]{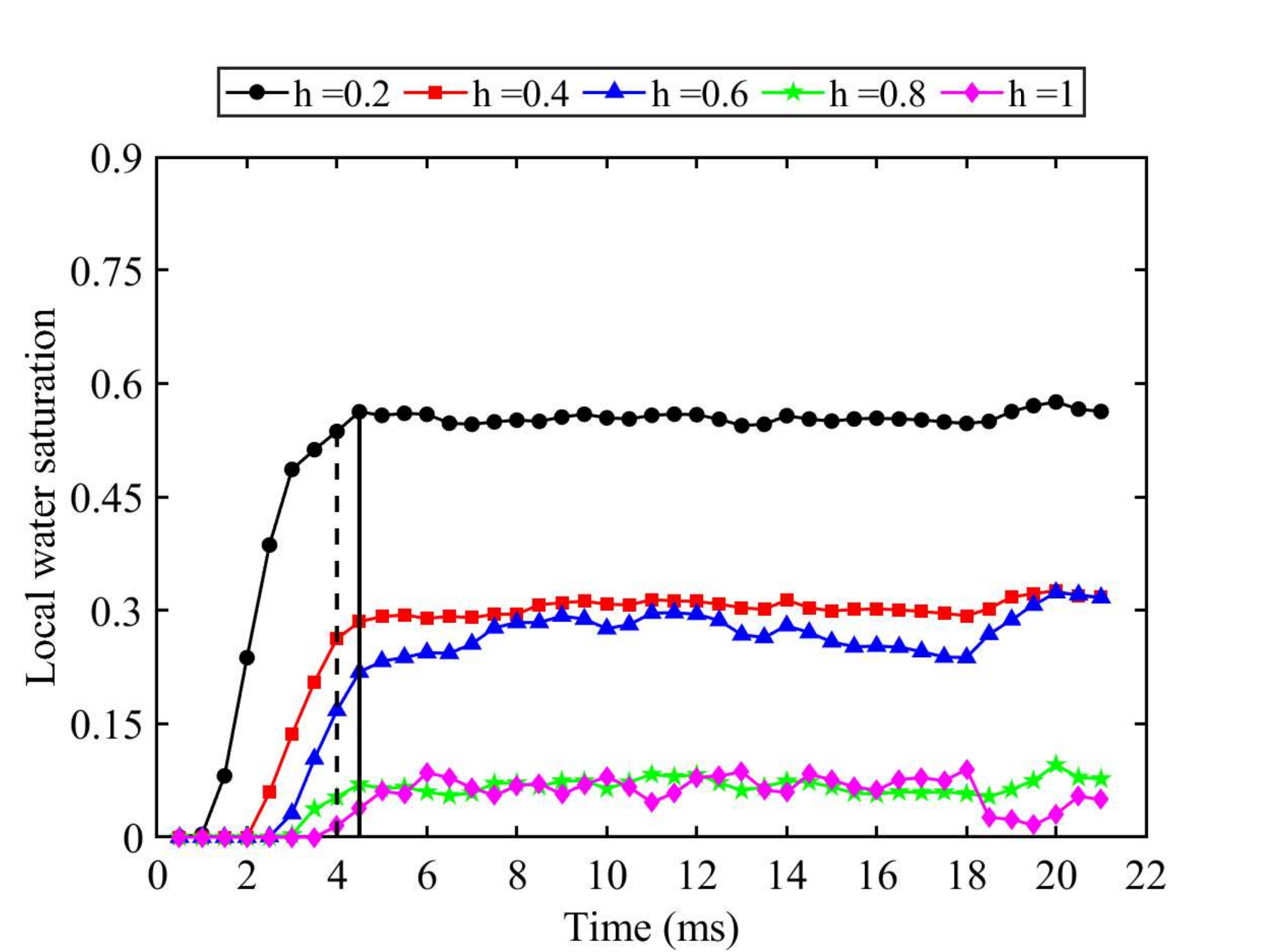}}
\caption{\label{fig:GDLLocalS}Time-varying local water saturation at five relative thicknesses (h = 0.2, 0.4, 0.6, 0.8, 1) extracted from X-Y cross-sections of the GDLs. (a) case A. (b) case B. (c) case C. The black dash line and solid line in each figure represent the breakthrough time and the stabilization beginning time.}
\end{figure}

The local water saturation along the relative thickness direction of three cases at t = 21 ms is shown in Fig. \ref{fig:Valiat2} (a), and a comparison with the study of Niu et al. \cite{niu2019two} is also presented. It is noted that 150 slice samples of the cross-sectional liquid distribution along the thickness are obtained to calculate the local water saturation in each case. The key difference of GDL geometric parameters in both studies is listed in Table \ref{tab:4}. The fiber diameter and porosity in the current work are larger than those in the reference study, which means the average pore radius of the GDLs is larger. However, the value of the contact angle in this research is larger, resulting in stronger hydrophobicity. Therefore, according to Eq. \ref{Equation9}, it is easy to explain the similar variation of local water saturation in the two studies owing to the opposite effects of these two parameters. We could see that the local water saturation in the two studies shows a good similarity in terms of trend and physical phenomenon in  Fig. \ref{fig:Valiat2}(a). It is conspicuous that the local water saturation varies from 1 at the GDL inlet boundary decreases to nearly 0 at the GDL/GC interface. The saturation of case A is higher than the other cases, while the saturation of case B is between that of case A and case C, and case C has the lowest local water saturation at almost all relative thickness locations. These results indicate case A has the most amount of water in the GDL. 

In order to better understand the difference between the two studies, the cross-sectional local porosity distribution along thickness in every case is also respectively presented in Fig. \ref{fig:Valiat2}(b-d). The variation of local porosity distribution in previous studies is almost uniform, which is apparently different from that in our studies, showing obvious variation around the bulk porosity value. In this regard, the fluctuation is thought to be realistic because there are no binder and PTFE microstructures being introduced to fill the irregular gaps between cylindrical fibers, and the porosity variation can be found in experimental research \cite{garcia2015effective}. In the connection region of two fiber layers in the reconstructed GDL, the pore area is bigger than that around it, thus an increase in the local porosity appears. In addition, compared with case B and case C, case A has more oscillation cycles due to a higher number of layers, as discussed in Section \ref{ReconstructionSection}. Although the local porosity of the three cases all fluctuates around the bulk porosity, the oscillation magnitude differs significantly. The porosity of case A varies in a smaller region of approximately 0.67-0.95 along the thickness direction, while the fluctuation range of case C is the largest, about 0.6-0.98.

The results indicate that the uniformity of local porosity decides the declining rate of local water saturation through the GDL thickness. The local water saturation of the previous study as well as case A and case B has a gradual decrease but case B shows a step-down trend, and the local water saturation near the inlet and outlet of the GDL decreases rapidly. Besides, a small variation of local water saturation is seen in all three cases because the local porosity changes the in-plane water flow area. The large pore area allows more liquid to pass, resulting in larger local water saturation. Thus the oscillation inevitably leads to the small oscillation of local water saturation along the thickness direction. The local water saturation of case C tends to fluctuate more. A larger variation of porosity will lead to a large change in local water saturation. Moreover, the \ref{LocalSAlongThickInDiffTimes} shows the local water saturation at several moments along the thickness for all three cases. It is found that all cases have a quick change of local water saturation at every position before the water breakthrough. After the breakthrough, all cases mainly have variation in the middle of the GDLs compared with that at the later time steps. However, the saturation of case B has a very small change in a thicker area, whereas that of case A has a big change in a larger thickness region.

\begin{table}[h]
\centering
\scriptsize
\caption{\label{tab:4} Comparison of GDL size parameters between the two studies.}
\begin{tabular}{lllllllllll}
\hline
 &&   Size  &&Fiber diameter &&  Contact angle && Average porosity \\\hline
 Case A && $1000\times1000\times300$ $\mu m$  && 10  $\mu m$   && $150^\circ$ && 0.81\\ 
Case B && $1000\times1000\times300$ $\mu m$  && 15  $\mu m$   && $150^\circ$ && 0.81\\ 
Case C && $1000\times1000\times300$ $\mu m$  && 20  $\mu m$   && $150^\circ$ && 0.81\\ 
Previous study \cite{niu2019two} && $800\times800\times192$ $\mu m$   &&   8 $\mu m$ && $109^\circ$ && 0.73 \\
\hline
\end{tabular}
\end{table}

\begin{figure}[H]
\subfigure[]{
\centering
\includegraphics[width=0.49\textwidth]{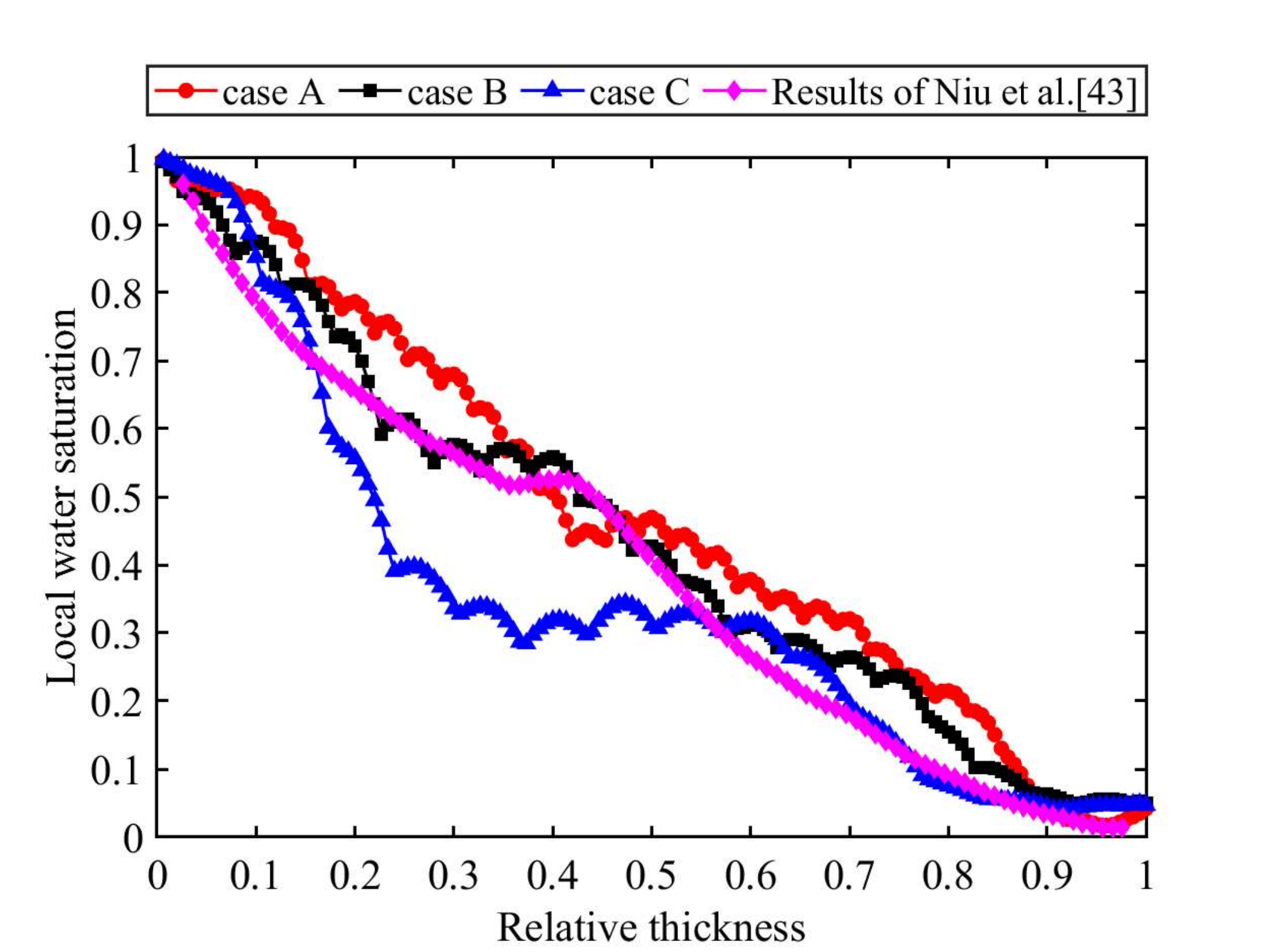}
}
\subfigure[]{
\centering
\includegraphics[width=0.49\textwidth]{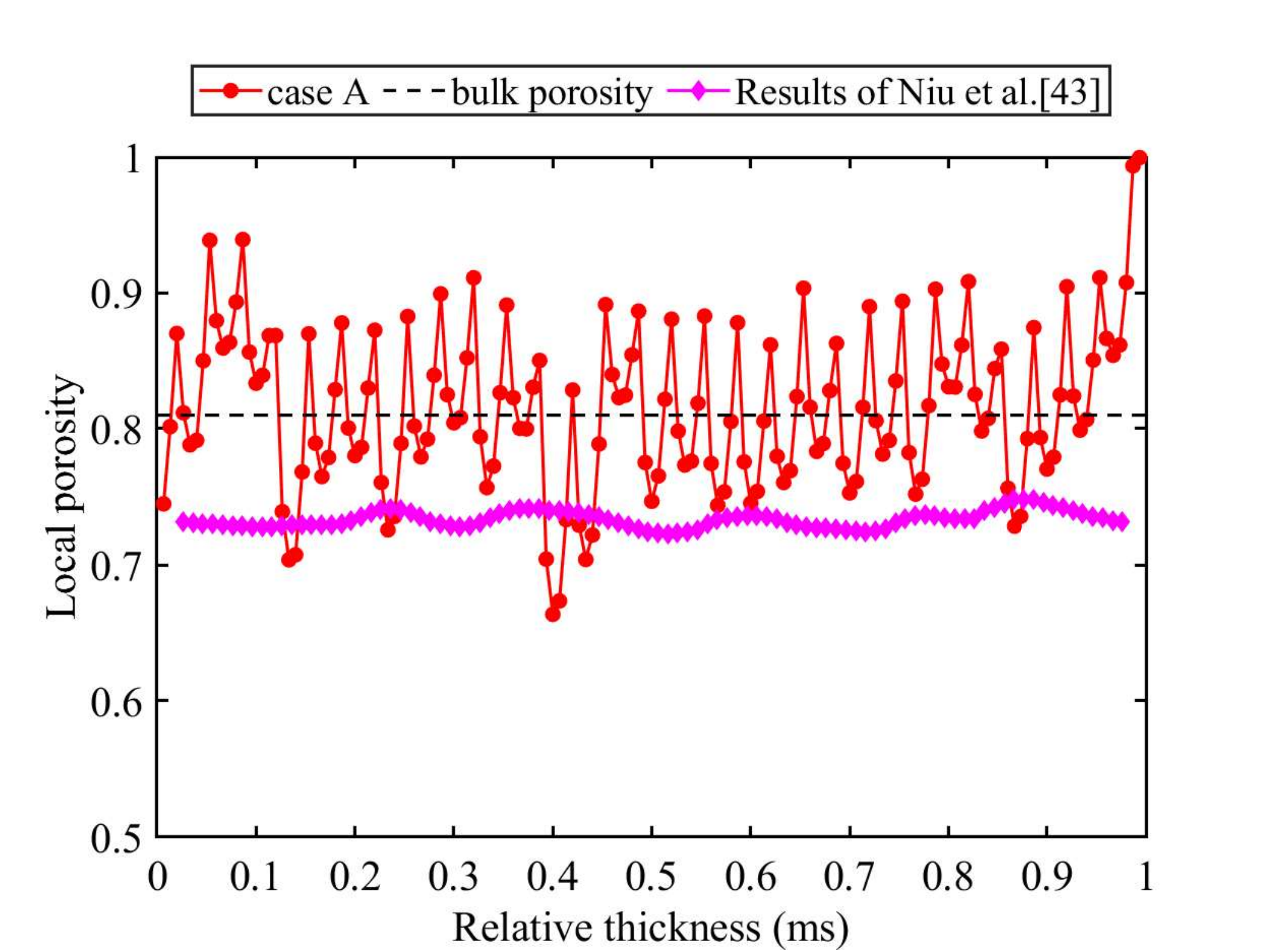}
}

\subfigure[]{
\centering
\includegraphics[width=0.49\textwidth]{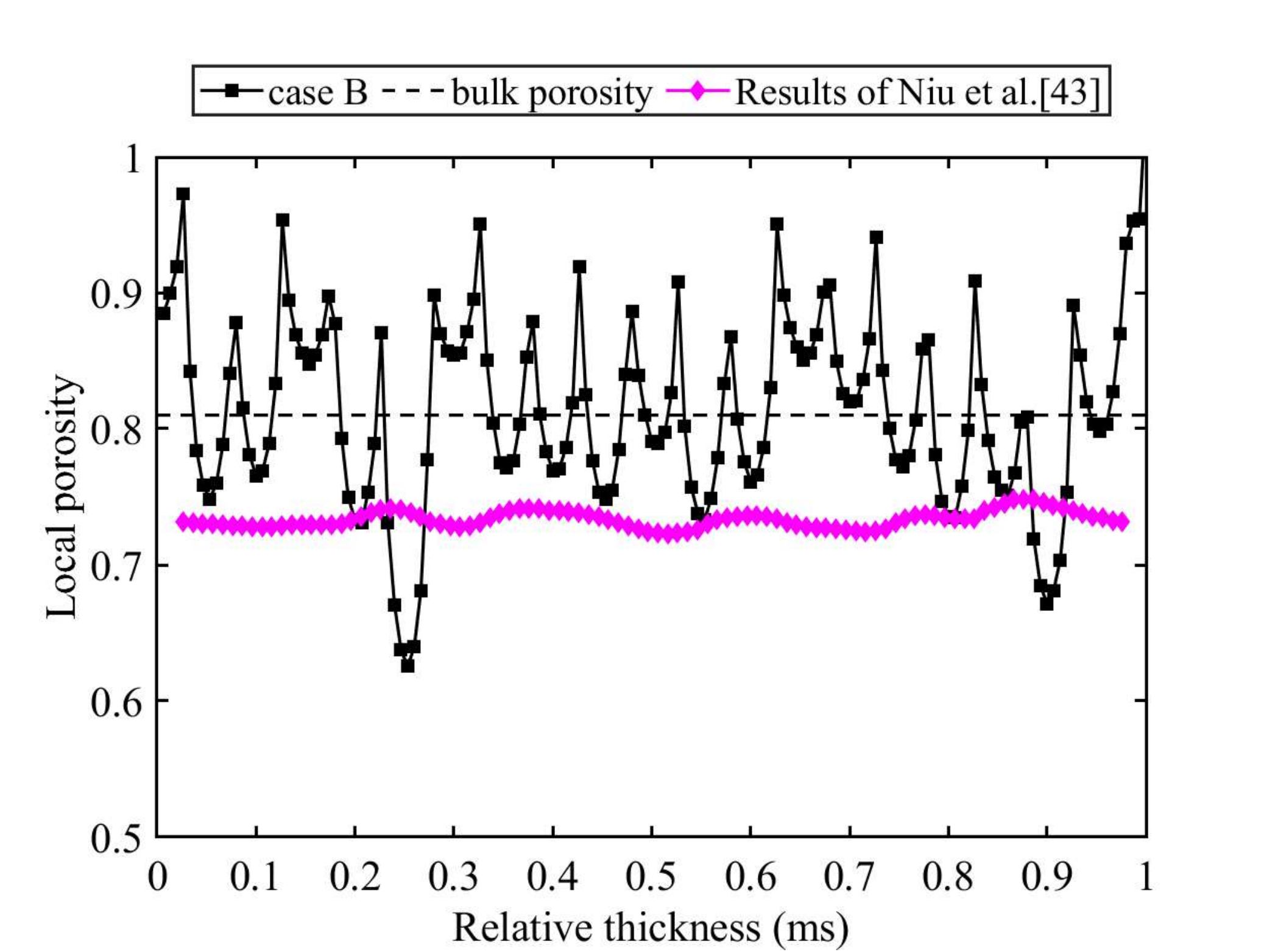}
}
\subfigure[]{
\centering
\includegraphics[width=0.49\textwidth]{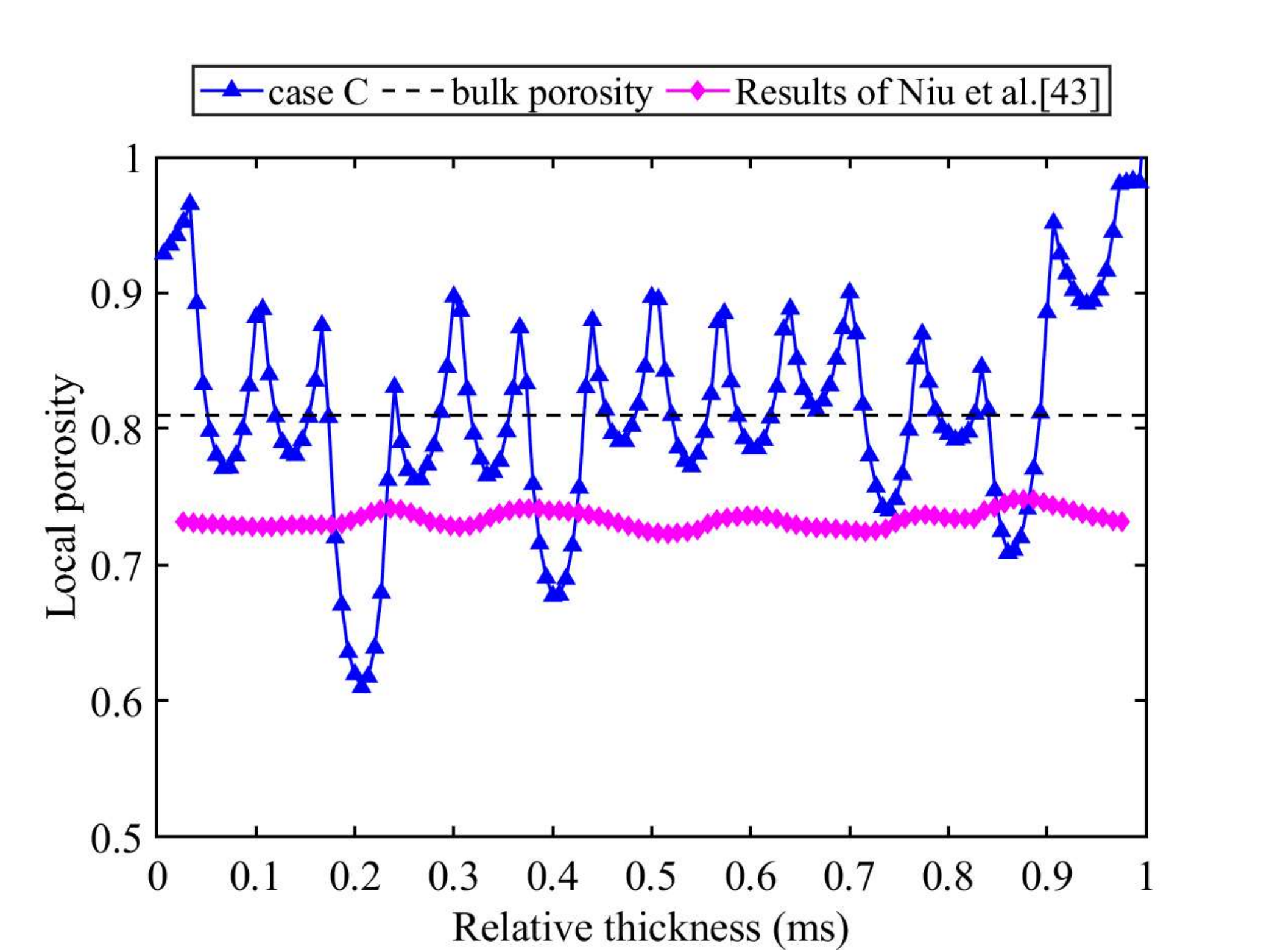}
}
\caption{\label{fig:Valiat2} Comparison of the local liquid water saturation and local porosity in the through-plane direction of GDLs between the present results and previous study \cite{niu2019two}. (a) The local liquid water saturation of case A (red), case B (black), case C (blue), and results of Niu et al. \cite{niu2019two} (pink); (b-d) The local porosity, in which the dash line means the bulk porosity of 0.81 and the pink line means the porosity in Niu et al. \cite{niu2019two}.}
\end{figure}

% \begin{figure}[H]
% \subfigure[]{
% \centering
% \includegraphics[width=0.49\textwidth, trim=10 225 40 240,clip]{Paper1Reverse1Pictures/localwatersaturationComparison3Cases1.pdf}
% }
% \subfigure[]{
% \centering
% \includegraphics[width=0.49\textwidth, trim=10 225 40 240,clip]{Paper1Reverse1Pictures/localporosityComparison3Cases1.pdf}
% }
% \caption{\label{fig:Valiat2} (a-b) Comparison of the local liquid water saturation and local porosity in the through-plane direction between the three GDLs in present study.}
% \end{figure}

\subsection{Total water saturation in GCs and GDLs}\label{TotalSGDLGCs}
The water coverage ratio in channel walls and the local water saturation in GDLs give us more detailed discoveries of the water behavior in GCs and GDLs. To obtain a more complete comparison between three cases in the total simulation domain, the time-varying total water saturation in both domains is shown in Fig. \ref{fig:TotalS}. It is found that the total water saturation in three GCs shows an increasing trend after the water breakthrough and the total water saturation in three GDLs rises first and then stabilizes. Besides, from case A to case C, the water saturation in the GCs increases sequentially at each moment before 18.5 ms. After 18.5 ms, the saturation value of case C starts to decrease due to a big droplet flowing out of the channel, as shown in Fig. \ref{fig:3Dwater3Cases}. In the later time, its same water saturation increase as that of case B mainly results from the same amount of water breakthrough to the GC from the corresponding GDL. Furthermore, case A always has the lowest total water saturation in the GC, and the saturation starts to increase from a late time of 6 ms due to the late breakthrough time.

During the increase of total water saturation in GDLs, the total water saturation in the GDL of case A is smaller than that of case B and case C. Furthermore, case C gets a lower stable saturation firstly, at around 0.3, followed by case B which is at an intermediate saturation of about 0.37. The results indicate that the GDL with a smaller fiber diameter leads to a slower water accumulation rate due to complex flow paths but has a high water saturation. As for the value of the stable total water saturation in the GDLs, they are similar to the previous results which are obtained only by simulating the water transport in the GDL \cite{niu2019two,lee2009pore}, lying in an approximate range of 0.3-0.45. Note that total water saturation values from Niu et al. \cite{niu2019two} are obtained by integrating the local water saturation along the GDL thickness.

% The water saturation in the GCs is several times smaller than that in the GDLs. Fig. \ref{fig:TotalS} (c) shows a conclusive comparison of the time-averaged total water saturation of three cases. The larger the water saturation in the GDL, the smaller the water saturation occurs in the GC, which indicates a balance of the water saturation when considering the design of GDLs.

\begin{figure}[H]
\centering
\subfigure[]{
\includegraphics[width=0.47\textwidth]{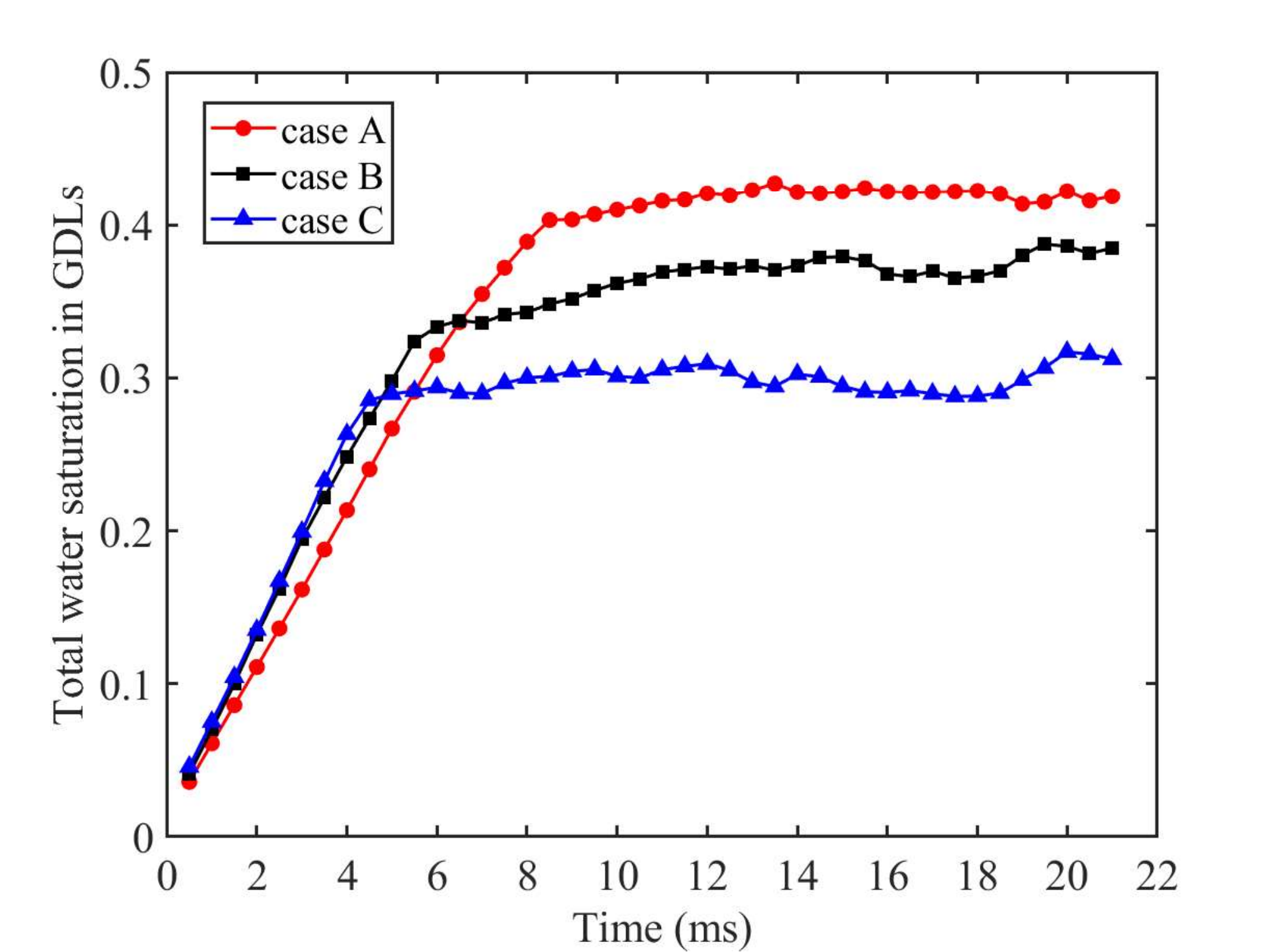}
}
\centering
\subfigure[]{
\includegraphics[width=0.47\textwidth]{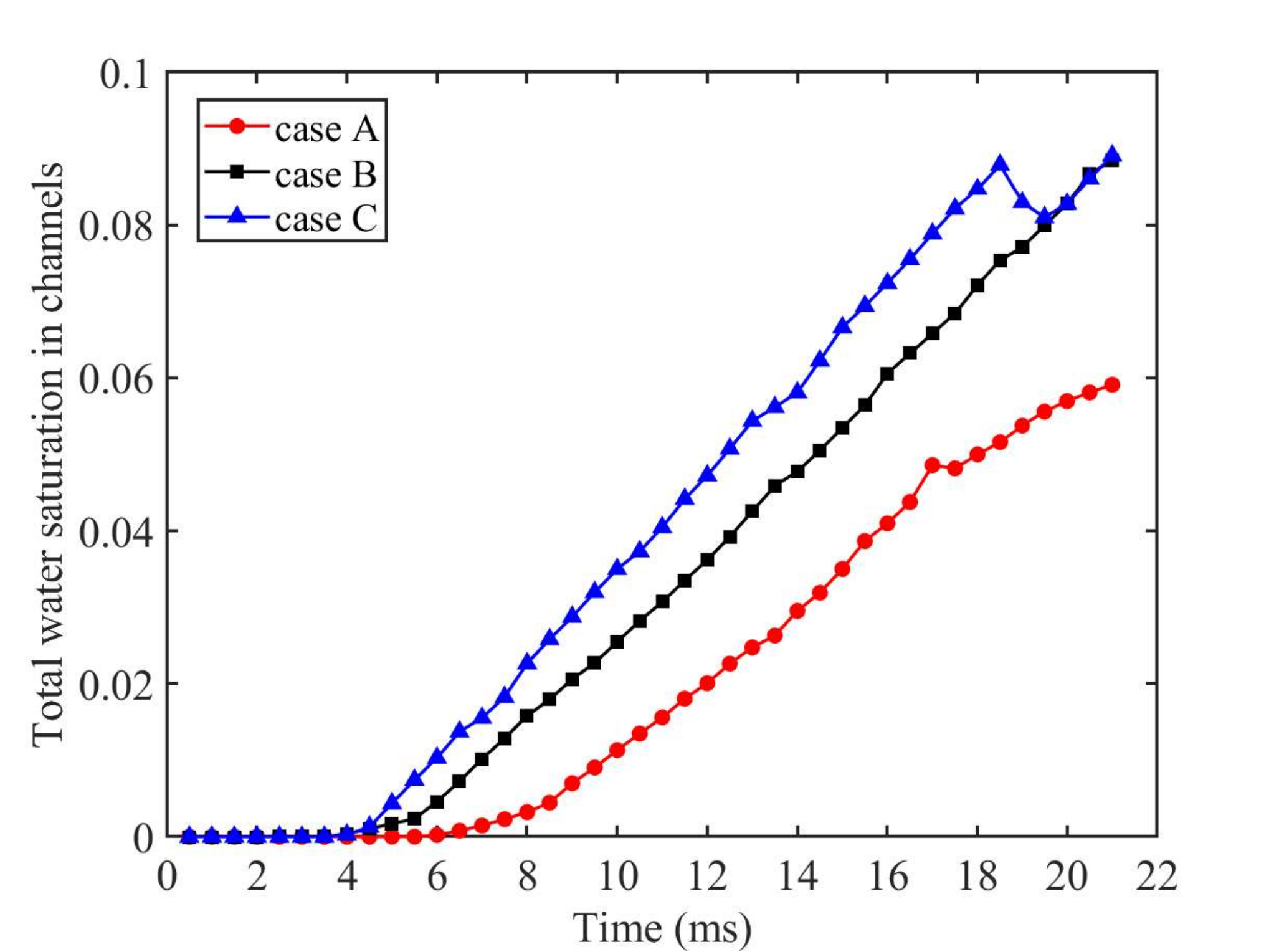}
}
% \centering
% \subfigure[]{
% \includegraphics[width=0.49\textwidth, trim=0 220 30 210,clip]{Paper1Reverse1Pictures/TotalSaZhuzi.pdf}
% }
\caption{\label{fig:TotalS} Comparison of the time-varying total liquid water saturation in GDLs and GCs for three cases. (a) In GDLs. (b) In GCs. Two figures show a total water saturation balance requirement between the GDL and GC.}
\end{figure}

\section{Conclusion}\label{Section4}
In the present study, detailed two-phase flow behavior in a "T-shape" connected GDL and GC domain in a PEMFC are investigated by adopting the VOF method. To study the effect of fiber diameter on water behavior, an in-house design procedure is used to stochastically reconstruct three different GDLs by varying the fiber diameter but keeping the GDL bulk porosity and geometric dimensions constant. Besides, the two-phase flow pattern evolution is also discussed. 

The general water dynamic transport evolution in both the GDL and GC as well as the difference between the three cases are qualitatively analyzed by time-varying 3D and 2D water distribution. The liquid water within the simulation domain experiences water invasion, water diffusion, void filling, and rising up in the GDL; droplets breakthrough at the GDL/GC interface; and accumulation, combination, attachment, and detachment in the GC region. It is found that a larger fiber diameter leads to a larger pore size distribution and simpler flow pathways, which reduce the microstructural resistance and decrease the capillary force, resulting in the water flowing out of the GDL faster. However, excessive accumulation of droplet flow in the corresponding GC is observed, forming large slugs. Besides, by means of reducing the fiber diameter, the liquid slug flow will become a water film and even be replaced by intermittent droplet flows. The physical forces in each region of the GDL and GC are also analyzed, and the liquid water flow in the GDL is found to be a capillary fingering flow type, due to the dominant surface tension force. The water distribution inside the GDL stabilizes after a certain time, while it shows instability with irregular oscillation around the GDL/GC interface.  Then, to get a further understanding of the water hydrodynamics, quantitative studies including the water coverage ratio at the GC walls as well as the local and total water saturation are conducted. Resulting of the huge difference in the wettability of the channel walls, the water coverage ratio at the GDL/GC connected surface is less than 5 $\%$, which is much smaller than the coverage ratio on the GC top and side walls. However, it shows irregular fluctuations within the small range, corresponding to that observed in the aforementioned 2D water distribution snapshots. The results show the hydrophilic walls in the channel can indeed avoid the accumulation of breakthrough water droplets on the surface of the GDL/GC, thus facilitating the discharge of water. As for the water saturation, the local and total water saturation in the GDLs increase quickly and then stabilize after the water breakthrough. Moreover, the local water saturation in three GDLs shows a decreasing trend along the thickness direction at any moment. An increasing trend in the GC total water saturation is found in our current study, which differs from the previous results \cite{zhang2022numerical}, showing that the water in the GC oscillates within a small range after a certain time. In this regard, increasing our simulation time could be considered in future research. In addition, a water saturation balance is observed in the connected GDL and GC, i.e., the more water saturation in the GDL, the less water saturation in the GC. Note that the case with the smallest fiber diameter of 10 $\upmu$m has the largest GDL total water saturation, with about 0.4, which is similar to the previous results obtained by only simulating GDL \cite{niu2019two,lee2009pore}. In our opinion, the question of selecting an appropriate fiber diameter still needs to be addressed in later studies.

In this study, heat transfer is not included, i.e., neither phase-change phenomenon nor electrochemical reactions are considered, but may be included in future extensions. In addition, the effect of fiber diameter on other GDL properties, such as permeability, diffusivity, thermal conductivity, and electrical conductivity, could be studied with the reconstructed GDLs. Finally, a comparison of experimental and stochastic reconstructed GDLs will be studied in future work. 

\section*{Acknowledgement}
The authors greatly appreciate the financial support of the Chinese Scholarship Council (grant number: 202006070174). 
Computer time was provided by the Swedish National Infrastructure for Computing (SNIC), partially funded by the Swedish Research
Council through grant agreement no. 2018-05973.

% Computer time is provided in part by the Swedish National Infrastructure for Computing (SNIC 2020/3-27)

\appendix 
\section{Stochastic reconstruction of the GDL}
\label{appendix:reconstruction}
As for the in-house reconstruction procedure, a geometry-based stochastic generation algorithm is developed and programmed in MATLAB R2020a. Due to the inconvenience of converting the generated structure to an STL file in MATLAB, COMSOL Multiphysics 5.6 software is linked with MATLAB, where the reconstruction program code is realized in MATLAB software, and the COMSOL is used to display the generated geometry in real-time, as shown in Fig. \ref{fig:appendix1}. Once the geometry is reconstructed and the set bulk porosity is reached, the STL format will be exported from the COMSOL software and used in OF7. Some assumptions are as follows: 
\begin{enumerate}
\item The carbon fibers are defined as long straight cylinders, and they randomly distribute in the reconstruction domain.
\item Along the GDL thickness direction, staggered overlapping fibers with penetration are allowed.
\item The diameter of fibers remains the same for one reconstruction.
\item The binder and PTFE microstructure is ignored in the fiber reconstruction process.
\end{enumerate}

\begin{figure}[H]
\centering
\includegraphics[width=0.5\textwidth]{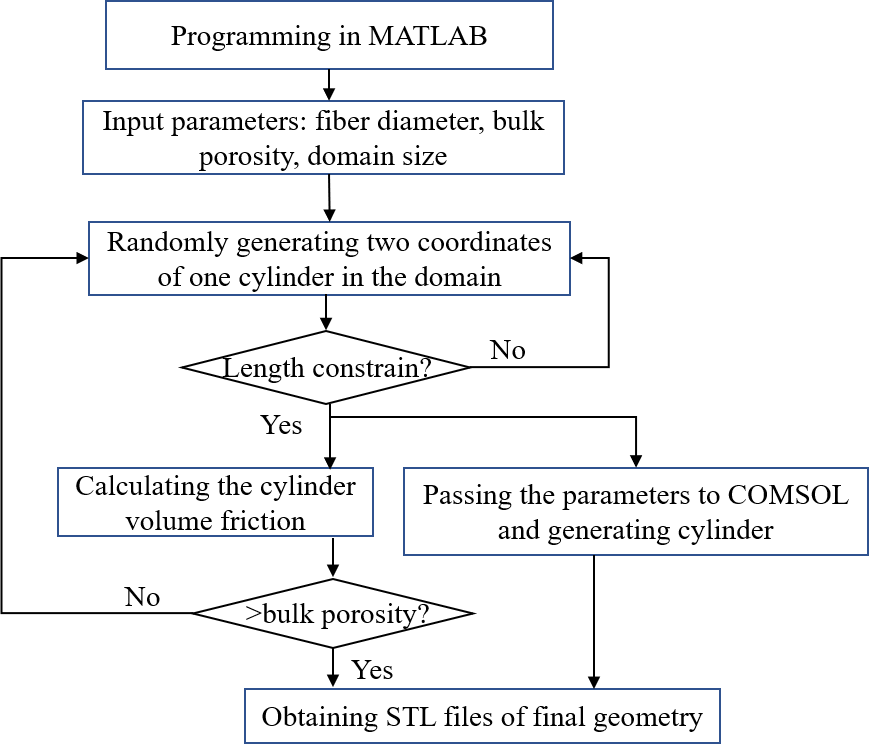}
\caption{\label{fig:appendix1} 3D numerical stochastic reconstruction of a PEMFC GDL. Geometric parameters such as fiber diameter, porosity, and geometric dimensions are used as input. MATLAB R2020a and COMSOL Multiphysics 5.6 are linked for the implementation of stochastic reconstruction code and visualization of 3D solid geometry, respectively.}
\end{figure}

\section{Local water saturation at different times along the GDL thickness}\label{LocalSAlongThickInDiffTimes}
\begin{figure}[H]
\centering
\subfigure[]{\includegraphics[width=0.45\textwidth]{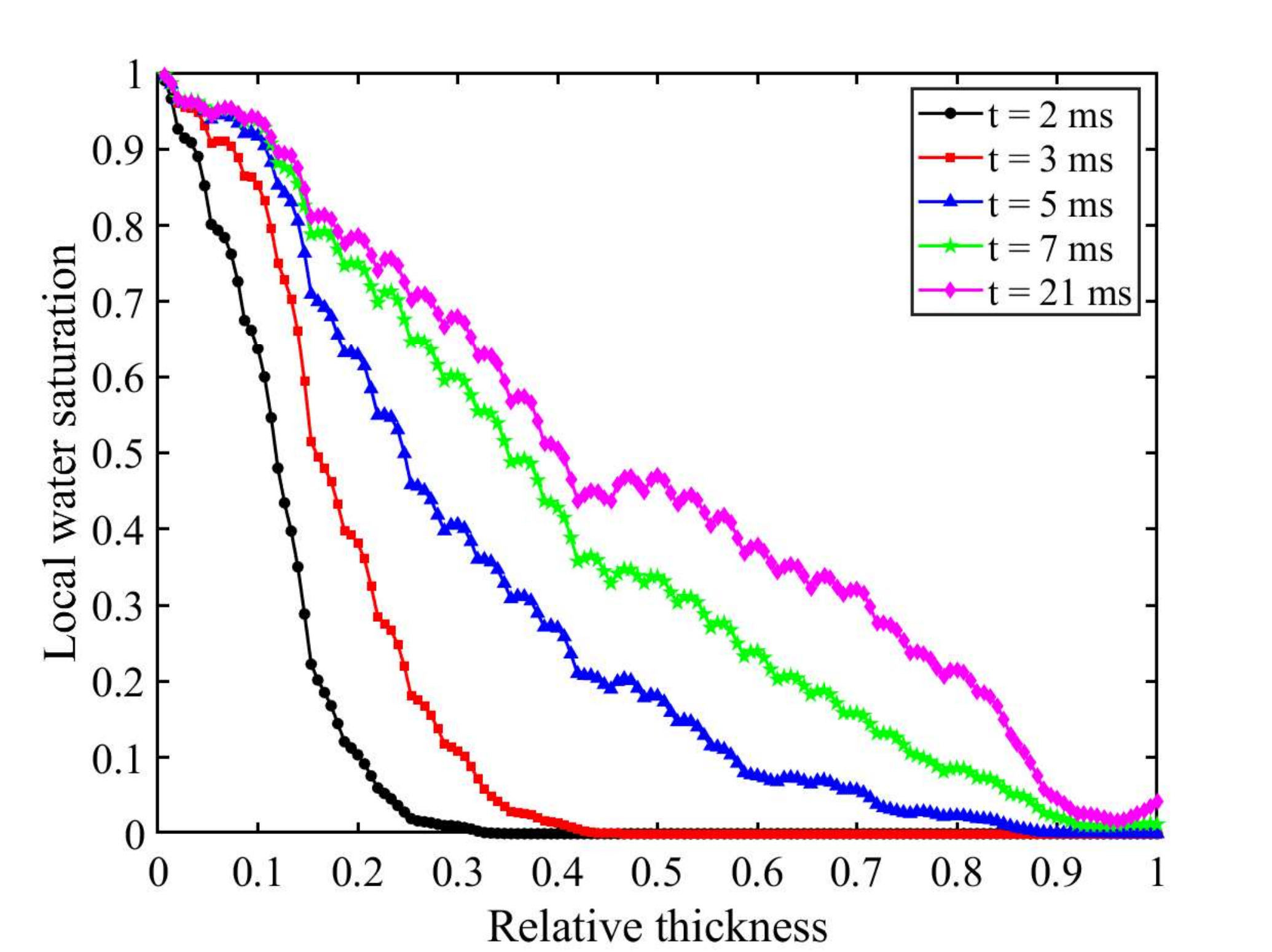}}
\centering
\subfigure[]{\includegraphics[width=0.45\textwidth]{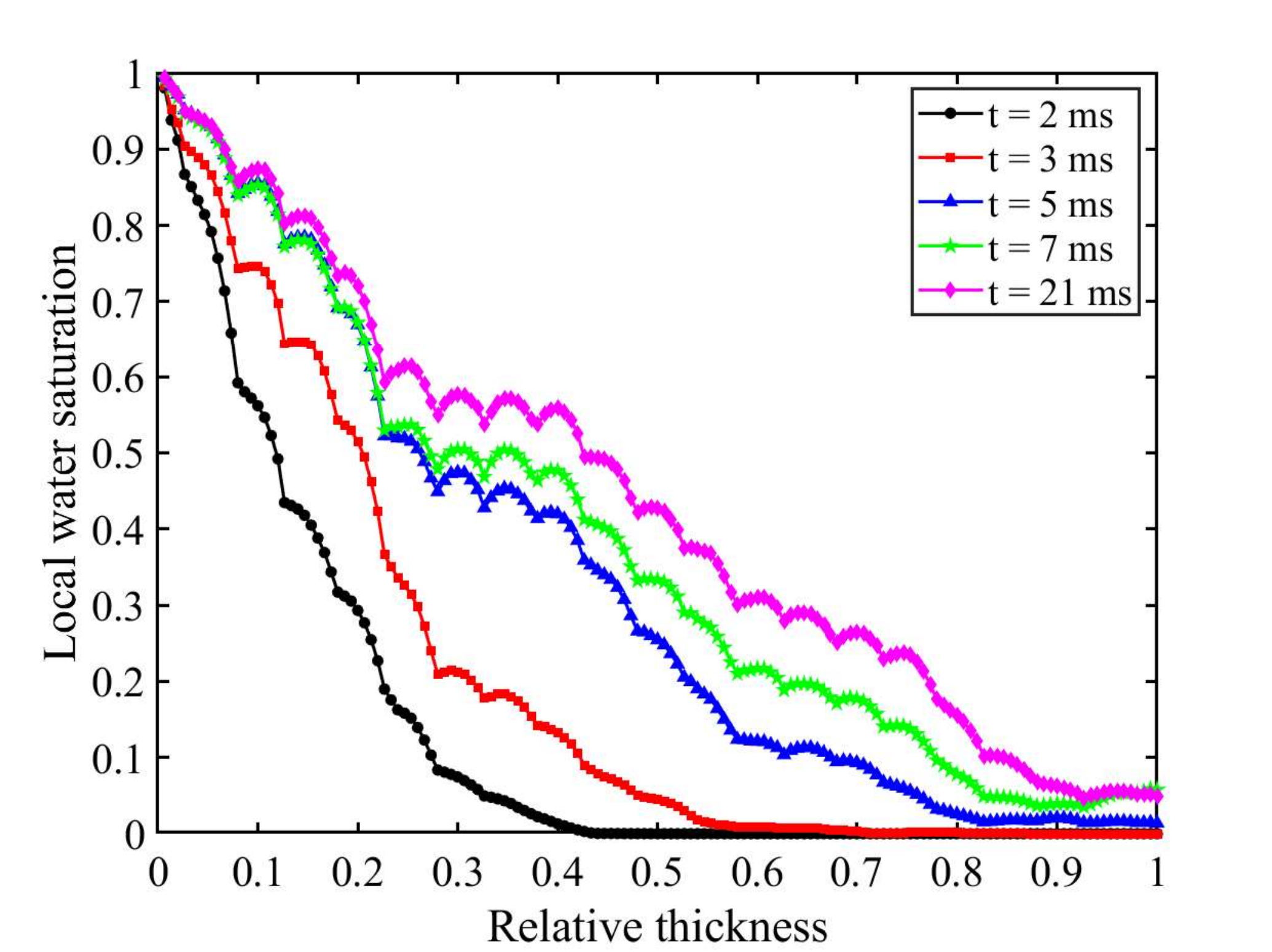}}

\centering
\subfigure[]{\includegraphics[width=0.45\textwidth]{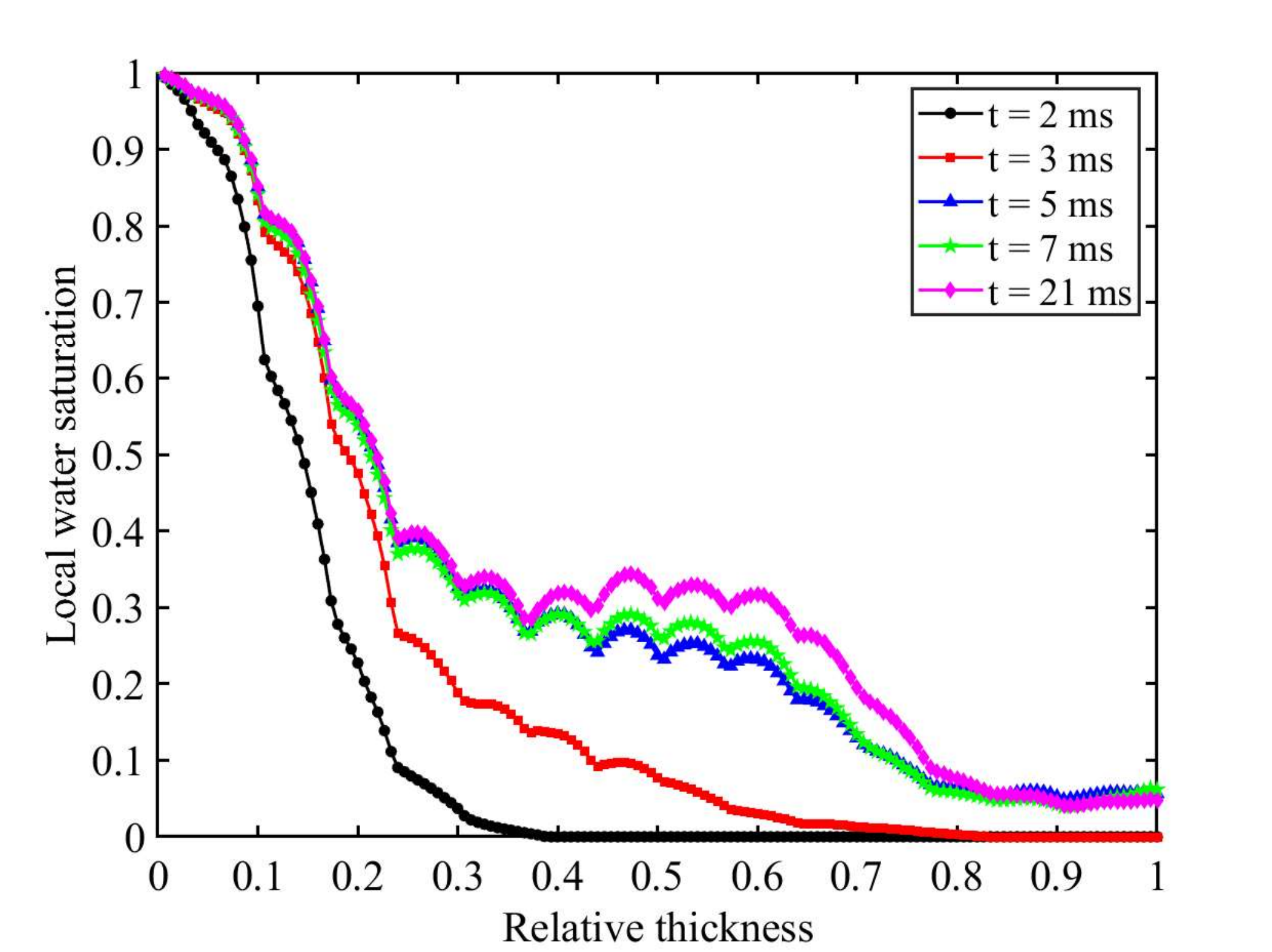}}
\caption{\label{fig:GDLLocalSAlongThick} Cross-section local water saturation in the GDL of three cases, varying along the relative thickness direction at different time steps, i.e., t = 2 ms, 3 ms, 5 ms, 7 ms, 21 ms. (a) case A. (b) case B. (c) case C.}
\end{figure}

\section{Supplementary data}
A video is attached as supplementary material to this article, showing liquid water transport through the GDL; droplet breakthrough at the GDL/GC interface; and water accumulation, combination, attachment, and detachment in the GC. 

\bibliographystyle{elsarticle-num.bst}

%\bibliographystyle{phys}
%\bibliography{sample}

\begin{thebibliography}{10}
\expandafter\ifx\csname url\endcsname\relax
  \def\url#1{\texttt{#1}}\fi
\expandafter\ifx\csname urlprefix\endcsname\relax\def\urlprefix{URL }\fi
\expandafter\ifx\csname href\endcsname\relax
  \def\href#1#2{#2} \def\path#1{#1}\fi

\bibitem{peighambardoust2010review}
S.~J. Peighambardoust, S.~Rowshanzamir, M.~Amjadi, Review of the proton
  exchange membranes for fuel cell applications, Int. J. Hydrog. Energy 35~(17)
  (2010) 9349--9384.

\bibitem{bao2006analysis}
C.~Bao, M.~Ouyang, B.~Yi, Analysis of water management in proton exchange
  membrane membrane fuel cells, Tsinghua Sci. Technol. 11~(1) (2006) 54--64.

\bibitem{pan2021review}
M.~Pan, C.~Pan, C.~Li, J.~Zhao, A review of membranes in proton exchange
  membrane fuel cells: Transport phenomena, performance and durability, Renew.
  Sust. Energ. Rev. 141 (2021) 110771.

\bibitem{r3}
Y.~Pan, H.~Wang, N.~P. Brandon, Gas diffusion layer degradation in proton
  exchange membrane fuel cells: Mechanisms, characterization techniques and
  modelling approaches, J. Power Sources 513 (2021) 230560.

\bibitem{zhu2021stochastically}
L.~Zhu, W.~Yang, L.~Xiao, H.~Zhang, X.~Gao, P.-C. Sui, Stochastically modeled
  gas diffusion layers: effects of binder and polytetrafluoroethylene on
  effective gas diffusivity, J. Electrochem. Soc. 168~(1) (2021) 014514.

\bibitem{r33}
M.~Bosomoiu, G.~Tsotridis, T.~Bednarek, Study of effective transport properties
  of fresh and aged gas diffusion layers, J. Power Sources 285 (2015) 568--579.

\bibitem{zhang2021microstructure}
H.~Zhang, L.~Zhu, H.~B. Harandi, K.~Duan, R.~Zeis, P.-C. Sui, P.-Y.~A. Chuang,
  Microstructure reconstruction of the gas diffusion layer and analyses of the
  anisotropic transport properties, Energy Convers. Manag. 241 (2021) 114293.

\bibitem{CHEN20218640}
Q.~Chen, Z.~Niu, H.~Li, K.~Jiao, Y.~Wang, Recent progress of gas diffusion
  layer in proton exchange membrane fuel cell: Two-phase flow and material
  properties, Int. J. Hydrog. Energy 46~(12) (2021) 8640--8671.

\bibitem{andersson2016review}
M.~Andersson, S.~Beale, M.~Espinoza, Z.~Wu, W.~Lehnert, A review of cell-scale
  multiphase flow modeling, including water management, in polymer electrolyte
  fuel cells, Appl. Energy 180 (2016) 757--778.

\bibitem{patel2019investigating}
V.~Patel, L.~Battrell, R.~Anderson, N.~Zhu, L.~Zhang, Investigating effect of
  different gas diffusion layers on water droplet characteristics for proton
  exchange membrane (pem) fuel cells, Int. J. Hydrog. Energy 44~(33) (2019)
  18340--18350.

\bibitem{BAO2021121608}
Z.~Bao, Y.~Li, X.~Zhou, F.~Gao, Q.~Du, K.~Jiao, Transport properties of gas
  diffusion layer of proton exchange membrane fuel cells: Effects of
  compression, Int. J. Heat Mass Transf. 178 (2021) 121608.

\bibitem{chen2021numerical}
G.~Chen, Q.~Xu, J.~Xuan, J.~Liu, Q.~Fu, W.~Shi, H.~Su, L.~Xing, Numerical study
  of inhomogeneous deformation of gas diffusion layers on proton exchange
  membrane fuel cells performance, J. Energy Storage 44 (2021) 103486.

\bibitem{guo2022pore}
L.~Guo, L.~Chen, R.~Zhang, M.~Peng, W.-Q. Tao, Pore-scale simulation of
  two-phase flow and oxygen reactive transport in gas diffusion layer of proton
  exchange membrane fuel cells: Effects of nonuniform wettability and porosity,
  Energy 253 (2022) 124101.

\bibitem{niu2018two}
Z.~Niu, Z.~Bao, J.~Wu, Y.~Wang, K.~Jiao, Two-phase flow in the
  mixed-wettability gas diffusion layer of proton exchange membrane fuel cells,
  Appl. Energy 232 (2018) 443--450.

\bibitem{sarkezi2022lattice}
P.~Sarkezi-Selsky, H.~Schmies, A.~Kube, A.~Latz, T.~Jahnke, Lattice boltzmann
  simulation of liquid water transport in gas diffusion layers of proton
  exchange membrane fuel cells: Parametric studies on capillary hysteresis, J.
  Power Sources 535 (2022) 231381.

\bibitem{zhang2022study}
S.~Zhang, S.~Xu, F.~Dong, Study on ice-melting performance of gradient gas
  diffusion layer in proton exchange membrane fuel cell, Int. J. Hydrog. Energy
  (2022).

\bibitem{jiao2021vapor}
D.~Jiao, K.~Jiao, Q.~Du, Vapor condensation in reconstructed gas diffusion
  layers of proton exchange membrane fuel cell, Int. J. Energy Res. 45~(3)
  (2021) 4466--4478.

\bibitem{SHI2021}
X.~Shi, D.~Jiao, Z.~Bao, K.~Jiao, W.~Chen, Z.~Liu, Liquid transport in gas
  diffusion layer of proton exchange membrane fuel cells: Effects of
  micro-porous layer cracks, Int. J. Hydrog. Energy 47~(9) (2022) 6247--6258.

\bibitem{zhang2022numerical}
L.~Zhang, S.~Liu, Z.~Wang, R.~Li, Q.~Zhang, Numerical simulation of two-phase
  flow in a multi-gas channel of a proton exchange membrane fuel cell, Int. J.
  Hydrog. Energy 47~(40) (2022) 17713--17736.

\bibitem{andersson2018modeling}
M.~Andersson, A.~Mularczyk, A.~Lamibrac, S.~Beale, J.~Eller, W.~Lehnert,
  F.~B{\"u}chi, Modeling and synchrotron imaging of droplet detachment in gas
  channels of polymer electrolyte fuel cells, J. Power Sources 404 (2018)
  159--171.

\bibitem{ANDERSSON201911088}
M.~Andersson, V.~Vukčević, S.~Zhang, Y.~Qi, H.~Jasak, S.~Beale, W.~Lehnert,
  Modeling of droplet detachment using dynamic contact angles in polymer
  electrolyte fuel cell gas channels, Int. J. Hydrog. Energy 44~(21) (2019)
  11088--11096.

\bibitem{ANDERSSON20182961}
M.~Andersson, S.~Beale, U.~Reimer, W.~Lehnert, D.~Stolten, Interface resolving
  two-phase flow simulations in gas channels relevant for polymer electrolyte
  fuel cells using the volume of fluid approach, Int. J. Hydrog. Energy 43~(5)
  (2018) 2961--2976.

\bibitem{Beale_2020}
S.~B. Beale, M.~Andersson, N.~Weber, H.~Marschall, W.~Lehnert, Combined
  two-phase co-flow and counter-flow in a gas channel/porous transport layer
  assembly, ECS Trans. 98~(9) (2020) 305--315.

\bibitem{qu2022proton}
E.~Qu, X.~Hao, M.~Xiao, D.~Han, S.~Huang, Z.~Huang, S.~Wang, Y.~Meng, Proton
  exchange membranes for high temperature proton exchange membrane fuel cells:
  Challenges and perspectives, J. Power Sources 533 (2022) 231386.

\bibitem{ashorynejad2019evaluation}
H.~R. Ashorynejad, K.~Javaherdeh, Evaluation of passive and active lattice
  boltzmann method for pem fuel cell modeling, Physica A 535 (2019) 121943.

\bibitem{ashorynejad2016investigation}
H.~R. Ashorynejad, K.~Javaherdeh, Investigation of a waveform cathode channel
  on the performance of a pem fuel cell by means of a pore-scale
  multi-component lattice boltzmann method, J. Taiwan Inst. Chem. Eng. 66
  (2016) 126--136.

\bibitem{mukherjee2020estimation}
M.~Mukherjee, C.~Bonnet, F.~Lapicque, Estimation of through-plane and in-plane
  gas permeability across gas diffusion layers (\makeuppercase{GDLs}):
  Comparison with equivalent permeability in bipolar plates and relation to
  fuel cell performance, Int. J. Hydrog. Energy 45~(24) (2020) 13428--13440.

\bibitem{CHEN20168550}
W.~Chen, F.~Jiang, Impact of ptfe content and distribution on liquid–gas flow
  in pemfc carbon paper gas distribution layer: 3d lattice boltzmann
  simulations, Int. J. Hydrog. Energy 41~(20) (2016) 8550--8562.

\bibitem{jeon2015effect}
D.~H. Jeon, H.~Kim, Effect of compression on water transport in gas diffusion
  layer of polymer electrolyte membrane fuel cell using lattice boltzmann
  method, J. Power Sources 294 (2015) 393--405.

\bibitem{nazemian2020impact}
M.~Nazemian, G.~Molaeimanesh, Impact of carbon paper structural parameters on
  the performance of a polymer electrolyte fuel cell cathode via lattice
  boltzmann method, Acta Mech. Sin. 36~(2) (2020) 367--380.

\bibitem{mangal2015experimental}
P.~Mangal, L.~M. Pant, N.~Carrigy, M.~Dumontier, V.~Zingan, S.~Mitra,
  M.~Secanell, Experimental study of mass transport in pemfcs: Through plane
  permeability and molecular diffusivity in \makeuppercase{GDLs}, Electrochim.
  Acta 167 (2015) 160--171.

\bibitem{schulz2007modeling}
V.~P. Schulz, J.~Becker, A.~Wiegmann, P.~P. Mukherjee, C.-Y. Wang, Modeling of
  two-phase behavior in the gas diffusion medium of pefcs via full morphology
  approach, J. Electrochem. Soc. 154~(4) (2007) B419.

\bibitem{gostick2006plane}
J.~T. Gostick, M.~W. Fowler, M.~D. Pritzker, M.~A. Ioannidis, L.~M. Behra,
  In-plane and through-plane gas permeability of carbon fiber electrode backing
  layers, J. Power Sources 162~(1) (2006) 228--238.

\bibitem{toetzke2014three}
C.~Toetzke, G.~Gaiselmann, M.~Osenberg, J.~Bohner, T.~Arlt, H.~Markoetter,
  A.~Hilger, F.~Wieder, A.~Kupsch, B.~R. M{\"u}ller, et~al., Three-dimensional
  study of compressed gas diffusion layers using synchrotron x-ray imaging, J.
  Power Sources 253 (2014) 123--131.

\bibitem{qiu2018electrical}
D.~Qiu, H.~Jan{\ss}en, L.~Peng, P.~Irmscher, X.~Lai, W.~Lehnert, Electrical
  resistance and microstructure of typical gas diffusion layers for proton
  exchange membrane fuel cell under compression, Appl. Energy 231 (2018)
  127--137.

\bibitem{hinebaugh2017stochastic}
J.~Hinebaugh, J.~Gostick, A.~Bazylak, Stochastic modeling of polymer
  electrolyte membrane fuel cell gas diffusion layers--part 2: A comprehensive
  substrate model with pore size distribution and heterogeneity effects, Int.
  J. Hydrog. Energy 42~(24) (2017) 15872--15886.

\bibitem{hirt1981volume}
C.~W. Hirt, B.~D. Nichols, Volume of fluid (\makeuppercase{VOF}) method for the
  dynamics of free boundaries, J. Comput. Phys. 39~(1) (1981) 201--225.

\bibitem{BRACKBILL1992335}
J.~Brackbill, D.~Kothe, C.~Zemach, A continuum method for modeling surface
  tension, J. Comput. Phys. 100~(2) (1992) 335--354.

\bibitem{jasak2009openfoam}
The \makeuppercase{OpenFOAM Foundation}, \url{http:://www.openfoam.org}; 2011,
  [accessed 3 August 2011].

\bibitem{DING2014469}
Y.~Ding, X.~Bi, D.~Wilkinson, Numerical investigation of the impact of
  two-phase flow maldistribution on pem fuel cell performance, Int. J. Hydrog.
  Energy 39~(1) (2014) 469--480.

\bibitem{DING20107278}
Y.~Ding, H.~Bi, D.~Wilkinson, Three-dimensional numerical simulation of water
  droplet emerging from a gas diffusion layer surface in micro-channels, J.
  Power Sources 195~(21) (2010) 7278--7288.

\bibitem{fluckiger2011investigation}
R.~Fl{\"u}ckiger, F.~Marone, M.~Stampanoni, A.~Wokaun, F.~N. B{\"u}chi,
  Investigation of liquid water in gas diffusion layers of polymer electrolyte
  fuel cells using x-ray tomographic microscopy, Electrochim. Acta 56~(5)
  (2011) 2254--2262.

\bibitem{niu2019two}
Z.~Niu, J.~Wu, Z.~Bao, Y.~Wang, Y.~Yin, K.~Jiao, Two-phase flow and oxygen
  transport in the perforated gas diffusion layer of proton exchange membrane
  fuel cell, Int. J. Heat Mass Transf. 139 (2019) 58--68.

\bibitem{lenormand1988numerical}
R.~Lenormand, E.~Touboul, C.~Zarcone, Numerical models and experiments on
  immiscible displacements in porous media, J. Fluid Mech. 189 (1988) 165--187.

\bibitem{garcia2015effective}
P.~A. Garc{\'\i}a-Salaberri, G.~Hwang, M.~Vera, A.~Z. Weber, J.~T. Gostick,
  Effective diffusivity in partially-saturated carbon-fiber gas diffusion
  layers: Effect of through-plane saturation distribution, Int. J. Heat Mass
  Transf. 86 (2015) 319--333.

\bibitem{lee2009pore}
K.-J. Lee, J.~H. Nam, C.-J. Kim, Pore-network analysis of two-phase water
  transport in gas diffusion layers of polymer electrolyte membrane fuel cells,
  Electrochim. Acta 54~(4) (2009) 1166--1176.

\end{thebibliography}
\end{document}